\documentclass[fleqn,usenatbib]{mnras}

\usepackage{newtxtext,newtxmath}
\usepackage[T1]{fontenc}
\usepackage{ae,aecompl}

\usepackage[english]{babel}
\usepackage{journals}
\usepackage[pdftex]{graphicx,color} 
\usepackage[latin1]{inputenc}
\usepackage{graphics} 
\usepackage{amsfonts} 
\usepackage{multicol} 
\usepackage{layout} 
\usepackage{layouts}
\usepackage{multirow}
\usepackage{float}
\usepackage{tabularx}
\usepackage{siunitx}
\usepackage[caption = false]{subfig}
\usepackage{graphicx}
\usepackage{ulem}

\defcitealias{niemiec2019}{N19}

\title[Satellite galaxy properties from the TNG simulation]{
Scatter in the satellite galaxy SHMR: fitting functions, scaling relations \& physical processes from the IllustrisTNG simulation}

\author[Niemiec et al.]
	{\parbox{\textwidth}{
	Anna Niemiec,$^{1,2}$\thanks{E-mail: \href{mailto:anna.niemiec@durham.ac.uk} {anna.niemiec@durham.ac.uk}} 
	Carlo Giocoli,$^{3,4}$
	Ethan Cohen,$^{5}$
	Mathilde Jauzac,$^{1,2,6,7}$
	Eric Jullo,$^{8}$
	Marceau Limousin$^{8}$
	\\
	}
	\vspace*{3pt}\\
$^{1}$Centre for Extragalactic Astronomy, Department of Physics, Durham University, Durham DH1 3LE, UK\\
$^{2}$Institute for Computational Cosmology, Durham University, South Road, Durham DH1 3LE, UK\\
$^{3}$INAF - Osservatorio di Astrofisica e Scienza dello Spazio di Bologna,  via Gobetti 93/3, I-40129 Bologna, Italy \\
$^{4}$INFN - Sezione di Bologna, viale Berti Pichat 6/2, I-40127 Bologna, Italy\\
$^{5}$University of Michigan, 1085 South University Ave, Ann Arbor, MI 48109, USA\\
$^{6}$Astrophysics Research Centre, University of KwaZulu-Natal, Westville Campus, Durban 4041, South Africa \\
$^{7}$School of Mathematics, Statistics \& Computer Science, University of KwaZulu-Natal, Westville Campus, Durban 4041, South Africa\\
$^{8}$Aix Marseille Univ, CNRS, CNES, LAM, 13388 Marseille, France \\
}

\date{Accepted XXX. Received YYY; in original form ZZZ}
\pubyear{2018}

\begin{document}

\label{firstpage}
\pagerange{\pageref{firstpage}--\pageref{lastpage}}
\maketitle


\begin{abstract}
The connection between galaxies and their dark matter haloes is often described with the Stellar-to-Halo Mass relation (SHMR).
Satellite galaxies in clusters follow a SHMR distinct from central galaxies because of the environmental
processes that they are subject to, and the variety of accretion histories leads to an important scatter in this
relation. In this work, we use the suite of  magneto-hydrodynamical simulations IllustrisTNG to study the
scatter in the satellite galaxy SHMR, and extract the parameters that can best allow to understand it.
Active galaxies, that represent a very small fraction of cluster galaxies, follow a very different relation than their passive 
counterparts, mainly because they were accreted much more recently.
For this latter population, we find that the distance to the cluster centre  is a good predictor
of variations in the SHMR, but some information on the galaxy orbital history, such as the distance of closest approach to 
the host centre, is an even better one, although it is in practice more difficult to measure. 
In addition, we found that galaxy compactness is also correlated with the SHMR, while the host cluster properties 
(mass and concentration, formation redshift, mass and size of BCG) do not play a significant role. 
We provide accurate fitting functions and scaling relations to the scientific community, useful to predict the subhalo mass given a set of observable parameters. Finally, we connect the scatter in the SHMR to the physical processes affecting galaxies in clusters, and how they impact the different satellite sub-populations. 

\end{abstract}


\section{Introduction}

In the standard model of cosmology, the formation of structures is driven by the dominant collision-less matter component: 
the dark matter (hereafter DM) \citep{white78,kauffmann1993,springel2010}. 
Galaxies are therefore believed to condense in the potential well of dark matter haloes that hierarchically merge forming 
larger structures through the cosmic time \citep{white1991,tormen1998,vandenbosch02,wechsler02,giocoli12b}. 
In this picture, the properties of galaxies are expected to be resolutely correlated with (at least) the mass of their host haloes. 
The relation that is most often used to quantify this galaxy-halo connection is the so called Stellar-to-Halo Mass Relation 
\citep[SHMR, see][for a review on the galaxy-halo connection]{wechsler2018}. 
Its shape has been well constrained using different observational probes such as 
gravitational lensing \citep[e.g.][]{mandelbaum2006,  hudson2015}, galaxy clustering combined with lensing \citep[e.g.][]{leauthaud2012,coupon2015} or galaxy group demographics \citep[e.g.][]{yang2012, rodriguez-puebla2015},
satellite kinematics \citep[e.g.][]{conroy2007, more2011},  or combining observations of galaxy properties 
with dark matter halo mass from N-body simulations 
\citep[for instance in a process called abundance matching, see][]{moster2010, behroozi2010,girelli20}. 
However, the scatter on the relation is still far from being fully understood, as different parameters can have an impact 
on this link, such as the star-formation rate of galaxies \citep{zehavi2011, rodriguez-puebla2015, mandelbaum2016} or 
their size \citep{somerville2018, sonnenfeld2019, posti19,huang2020,posti21}.

While this relation connects galaxies and their haloes on several orders of magnitudes, various physical processes 
will have an impact on it at different scales, from the low mass regime of dwarf galaxies \citep{read2017} 
to the extreme of giant galaxy clusters \citep{kravtsov2018}.
The very high mass-end presents a particular interest: galaxy clusters are the most massive gravitationally bound objects in 
the Universe \citep{tinker08,despali16}, with very high density of both dark and baryonic matter \citep{ettori09}. 
In addition, their long and complex accretion history bears the imprint of the formation of structures on their 
(baryonic and dark matter) mass distribution \citep[for a review, see][]{kravtsov2012}.  
Constraining the galaxy-halo connection in this environment is therefore an important probe of the nature of dark matter
\citep{despali19,lovell19a,lovell19b}, as well as of subtle baryonic processes, that shape the formation of structures. 
There are two possible approaches regarding the galaxy-halo connection in clusters: first, it is possible to constrain 
the overall mass distribution in clusters, combining different probes (such as for example lensing, X-ray, spectroscopy 
\citep{sereno13,bergamini19}) to distinguish the contribution from the different components (dark matter, gas, stars), 
and then connect this mass distribution to the underlying accretion history \citep{deboni16} and physical processes 
\citep[e.g.][]{richard2010, jauzac2015a}. 
But another approach that can be considered is to directly statistically examine the galaxy-connection for satellite galaxies 
and their host subhaloes, and therefore constrain the active mechanisms in the formation of galaxy clusters.

In this paper, we focus on this second approach. Galaxies are influenced by specific interactions while they infall into their host 
cluster-halo: on the baryonic side, interactions in the dense environment, such as ram-pressure stripping \citep{gunn&gott1972}, 
starvation/strangulation \citep{larson1980} or harassment \citep{moore1996,moore1998}, will tend to produce a population of passive 
galaxies. At the same time, tidal forces of the host can strip subhaloes from part of their dark matter
\citep{merritt1983,vandenbosch2005b}, while dynamical friction slowly make the galaxies sink towards the centre of their hosts 
\citep{ostriker&tremaine1975, binney&tremaine2008,nipoti18}. Measuring the strength and the impact of these different processes can open a window 
to the understanding of the nature of dark matter \citep{sirks2021} and the baryonic processes that govern galaxy evolution.

A powerful tool to measure observationally the Stellar-to-subHalo Mass Relation (SsHMR) for cluster galaxies is gravitational lensing. 
Galaxy-galaxy strong lensing can allow to put strong constraints on the subhalo mass of individual galaxies 
\citep{bergamini19,meneghetti20}, but such events are rare and such measurements are therefore more sensitive to intrinsic variability. 
The overall modelling of matter distribution in clusters includes a contribution from cluster members and can therefore be used to 
constrain subhalo masses \citep{grillo2015}, but it can be degenerate with the large scale mass distribution \citep{limousin2016}; and 
finally, galaxy-galaxy lensing in the weak regime allows to measure subhalo masses over stacked samples of galaxies 
\citep{li2015,sifon2015,niemiec2017,sifon2018}, but it requires large statistical data sets and a good understanding of the selection of 
galaxy samples and of the scatter therein.

In order to interpret observational results and disentangle the impact of dark matter and baryonic processes, 
it is important to parametrize scaling relations and recipes derived from state-of-the-art numerical simulation. 
Indeed, simulations allow to replicate the observed Universe, given some assumptions on initial conditions and physical processes. 
They represent a privileged tool to follow the evolution of galaxies over time, and study the different interactions they undergo.
Linking physical processes implemented in simulations with the observed Universe is not straightforward, 
but it allows to "de-project" in time and space the 2D picture of the sky that is the basis for all observational analyses. 
In addition, simulations can also allow to improve and drive observational studies, 
for instance by revealing new physical parameters that can trace some physical interactions. 

While these numerical techniques have allowed to model the gravitational evolution of galaxies, 
and the large scale matter distribution in the Universe, under the cold dark matter paradigm for decades 
\citep[see for instance][]{holmberg1941, press&schechter1974, springel2005, klypin2011}, 
the life and evolution of the baryonic component of galaxies remains more demanding to simulate. 
Indeed, it depends on many complex physical processes, acting on a variety of scales. 
Two main techniques have been developed in the past years: semi-analytical models (hereafter SAMs) and full hydrodynamical simulations.
SAMs \citep[e.g][]{white1991, kauffmann1993, delucia2007, somerville2008b, guo2010, lacey2016} rely principally 
on dark matter simulation merger-trees, that are then populated with seed galaxies, followed during various merging events 
along the cosmic time. These galaxies  evolve following analytical prescriptions, motivated by models derived from 
a combination of  theory and observations. 
The advantage of this approach is that it has relatively low computational cost and has proven quite successful in recovering many 
statistical properties of galaxies such as the stellar mass function \citep{guo2015}. 
However, it does not directly account for interactions between the baryonic and dark matter components, that can have a non-negligible 
impact, in particular in high-density environments \citep{dolag2009, haggar2021, bahe2021}. On the other hand, 
hydrodynamical simulations \citep[for a review see][]{vogelsberger2020} model galaxy formation processes by coupling gravity 
with gas physics, and thus reproduce the co-evolution of dark and baryonic matter in a more realistic way. 
However, they remain much more demanding in terms of computational power, which still limits their volume: 
the largest hydrodynamical simulation boxes such as IllustrisTNG now reach a few 100\,Mpc side length, 
while dark matter only universes have been simulated in boxes  of up to a few Gpc 
size \citep[eg. the Big MultiDark simulation, see][]{klypin2016}.
It is important to notice that both approaches are constructed and tailored to statistically reproduce global 
observables of galaxies and clusters. 

N-body and hydrodynamic simulations represent a particularly interesting tool to quantify the physical processes that influence
the properties of satellite galaxies, with respect to their central/field counterparts.  
Gravitational interactions, such as tidal stripping by the gravitational potential of the host 
\citep{vandenbosch2005b,giocoli08b}, but also by other subhaloes \citep{knebe2006}, create a decrease 
of the subhaloes dark matter mass that starts well outside the virial radius of the cluster \citep{behroozi2014}, 
and represent the main driver in the total subhalo mass evolution.
While various studies agree that satellite galaxies are mostly quenched by the cluster environment, 
the exact time scales and contributions of the different processes (tidal/ram-pressure stripping, harassment, strangulation/starvation, 
etc.) is still to be precisely quantified \citep[e.g][]{wetzel2013, jaffe2015, lotz2018, tremmel2019}.  
As demonstrated in \citet{donnari2021a}, $30\%$ of all quenched galaxies in massive groups and clusters at $z=0$ 
were already before infall due to internal quenching or interactions within smaller groups. 
Tidal stripping can also affect the stellar component of satellite galaxies, but only if most of the dark matter has 
already been stripped \citep{smith2016}.
The combination of these different mechanisms leads to a shift in the SHMR measured for satellite galaxies in clusters, 
as compared to the same relation for central/field galaxies \citep{neistein2011,rodriguez-puebla2012, rodriguez-puebla2013, reddick2013}. 
Particularly, in \citet{niemiec2019} \citepalias[hereafter][]{niemiec2019}, we measured this shift, 
and quantified the contribution of the different process (dark matter stripping, star formation quenching) to it. 
\citet{engler2021} led a similar study in the TNG simulation, and found that similar processes affect satellite galaxies 
not only in clusters, but also in groups with $M_{200} \leq 10^{12}M_{\odot}$. 
\citet{donnari2021a} analysed in details quenching mechanisms for satellite galaxies in groups and clusters, 
and found that low mass ($M_{\star} < 10^{10}M_{\odot}$) galaxies are mainly quenched by environmental interactions, 
while more massive galaxies are more subject to internal quenching. 
However, the scatter in the SsHMR is quite high \citep[even more than for central galaxies, see][]{rodriguez-puebla2013}, which shows that not all galaxies are affected to the same degree 
by the different types of interactions.
\citet{rhee2017} showed that considering the position of satellite galaxies in a phase-space diagram can help to partly 
understand this scatter. 
They also demonstrated that interactions within groups prior to the infall into the final host cluster, 
which is know as preprocessing \citep{mcgee2009, bahe2013, hou2014}, also affects the stellar and dark matter 
components of satellite galaxies, contributing to the SsHMR scatter \citep[see also][]{joshi2019}.
The relatively high scatter in the SsHMR has two important consequences: 
(i) observationally constraining the relation with precision can be difficult, and 
(ii) populating dark matter subhaloes in clusters with galaxies in N-body simulation can be imprecise 
if only using the subhalo mass.

In this paper, we use the publicly available state-of-the-art magneto-hydrodynamical simulation 
IllustrisTNG \citep{springel2018, pillepich2018a, naiman2018, marinacci2018, nelson2018a} 
to further study, model and interpret the scatter in the satellite galaxies SsHMR, and help improving the link between 
observations and simulations. 
We consider the problem as twofold: on one hand, the SsHMR can be considered from an "observational" point of view, 
meaning that given a set of observable parameters describing the satellite galaxy population, such as the stellar mass or the 
star-formation rate, it can be useful to predict the subhalo mass distribution, and determine which observational parameters have the 
most impact on it, and are therefore most correlated with the scatter in the SsHMR. This approach provides a useful comparison point for 
planning and interpreting observational studies. On the other hand, the link between observations and simulations can be considered from 
a theoretical point of view, and reducing the scatter in the constrained SsHMR can lead to an improved accuracy when populating N-body 
simulation with galaxies, taking into account for instance the orbital history of the subhaloes. To complement these two approaches, 
simulations also allow to examine the time evolution of subhaloes and their stellar counterpart, therefore linking the SsHMR and its 
scatter to physical mechanisms. 

This paper is structured as follow: 
In Sect.~\ref{sec:data}, we describe the TNG simulation that we use in our analysis, 
and the cluster/satellite galaxy selections that we apply. 
In Sect.~\ref{sec:shmr_obs}, we give some fitting functions that predict subhalo masses from stellar masses, 
and explore additional observable parameters that can improve the prediction. 
In Sect.~\ref{sec:shmr_simu}, we investigate the opposite approach, giving predictions for galaxy stellar masses 
as a function of subhalo masses, and other parameters that can be typically extracted from simulations, 
such as some proxy for the subhalo orbital history. 
Finally we link in Sect.~\ref{sec:time_evol} the measured SsHMRs to the physical process that take place in clusters, 
and discuss in Sect.~\ref{sec:discussion} the main differences that we observe with respect to the analysis on 
Illustris presented in \citet{niemiec2019} \citepalias[hereafter][]{niemiec2019}, and the possible impact of numerical resolution.

The cosmology used in this paper is identical to that used in the IllustrisTNG simulation, a flat $\Lambda$CDM 
universe consistent with the Planck 2015 results \citep[][$\Omega_{\rm{m,0}}=0.3089$, 
$\Omega_{\Lambda, 0} = 0.6911$, $\Omega_{\rm{b, 0}} = 0.0486$, $\sigma_8 = 0.8159, n_{\mathrm{s}} = 0.9667$ 
and $H_0 = 67.74 \rm{km\,s}^{-1}$]{planck2015}. 
The notation log() refers to the base 10 logarithm.

\section{Data}
\label{sec:data}
\subsection{The IllustrisTNG simulations}

IllustrisTNG  is a series of cosmological magneto-hydrodynamical simulations 
\citep{springel2018, pillepich2018a, naiman2018, marinacci2018, nelson2018a}, that represents an upgrade 
of the original Illustris runs \citep{genel2014, vogelsberger2014a, vogelsberger2014b}.
It models the coupled evolution of dark matter and gas dynamics using the quasi-Lagrangian code 
\textsc{arepo} \citep{springel2010}. 
In addition to the gravitational interaction and magneto-hydrodynamical evolution of the gas, 
it includes a galaxy evolution model with subgrid physical processes implemented, 
such as gas radiative cooling and heating, star formation \citep[following a Chabrier Initial stellar Mass Function,][]{chabrier2003} and evolution 
(and ensuing chemical enrichment of the environment), formation, evolution and feedback from super-massive black holes (SMBHs). 

The TNG simulations represent a improvement of the pre-existing Illustris runs, with some new developments included to introduce 
new physical processes, as well as relieve some tensions with observations 
\citep[such as insufficient quenching of star-formation, 
see][]{nelson2015}. 
In addition to various upgrades in the numerical methods and inclusion of magneto-hydrodynamics, 
the galaxy evolution model has also been refined, in three main specific areas: the growth and feedback of SMBHs, 
the modelling of galactic winds and of stellar evolution and gas enrichment  \citep{pillepich2018a}. 
Particularly, we will examine how this impacts the evolution of satellite galaxies in clusters in Sect.~\ref{sec:diffs_ill}.

Varius runs of the simulations are available, representing different simulation volumes and mass resolutions: 
the reference galaxy sample used in this paper is taken from the largest volume -- lower resolution run, TNG300, 
in order to obtain a maximal number of rare galaxy clusters, with the largest mass range (in TNG300, $M_{200}^{\mathrm{max}} = 
10^{15}h^{-1}M_{\odot}$ at $z = 0$). The TNG300 simulation box is $\sim300$~Mpc side length, with dark matter particle mass 
$m_{\mathrm{DM}} = 4\times 10^7 h^{-1}M_{\odot}$, average gas cell mass $m_{\mathrm{gas}} = 7.5\times 10^6 h^{-1}M_{\odot}$ 
and gravitational softening length $\epsilon_{\mathrm{DM}} = 1 h^{-1}\mathrm{kpc}$ at $z=0$.  

To examine the impact on our results of numerical effects, such as potentially unresolved galaxies/unconverged subhaloes,  we take 
advantage in some parts of the analysis of the higher resolution (but lower volume) simulation run: the TNG100 ($m_{\mathrm{DM}} = 5 \times 10^6 h^{-1} M_{\odot},~ m_{\mathrm{gas}} = 9.5 \times 10^5 h^{-1} M_{\odot},~\epsilon_{\mathrm{DM}} = 0.5h^{-1}\mathrm{kpc}$). 
We do not extract our fiducial galaxy sample from this simulation run as it contains a smaller number of galaxy clusters, 
and does not contain as massive haloes as the TNG300 run ($M_{200}^{\mathrm{max}} = 2.6 \times 10^{14} h^{-1} M_{\odot}$). 
In addition, for each run,  lower resolution versions of the simulation are available. 
For our fiducial sample based on TNG300, we only use the most resolved version, TNG300 $\equiv$ TNG300-1, 
correcting the galaxy stellar masses (see Sect~\ref{sec:sample}) using both the most resolved version, TNG100-1, and a lower resolution 
version, TNG100-2 ($m_{\mathrm{DM}} = 4 \times 10^7 h^{-1}M_{\odot},~ m_{\mathrm{gas}} = 7.5 \times 10^6 h^{-1} M_{\odot},~\epsilon_{\mathrm{DM}} = 
1 h^{-1}\mathrm{kpc}$), which has the same resolution as the TNG300-1 run. For the rest of the analysis, we remind the reader that 
TNG100~$\equiv$~TNG100-1.

In this paper, we use the publicly available group catalogue that was extracted from the simulation with a Friend-of-Friend algorithm 
with linking length $b=0.2$, as well as the available subhalo/galaxy catalogue that was obtained with \textsc{Subfind} 
\citep{springel2001, dolag2009}. 
We describe in Sect.~\ref{sec:sample} the associated quantities that are used in our analysis. 
To trace  the orbital and mass evolution of galaxies and subhaloes, 
we take advantage of the available merger trees that were extracted with the \textsc{SubLink} algorithm \citep{rodriguez-gomez2015}.

\subsection{Clusters and satellites in TNG}
\label{sec:sample}

To study the properties of satellite galaxies in clusters, we first select a sample of cluster-like haloes in the 
TNG300 simulation at redshift zero. 
These are defined following a simple mass selection, considering all systems with $M_{200} > 9 \times 10^{13} h^{-1}M_{\odot}$, 
resulting in a sample of 177 clusters. 
The quantities associated to the cluster sample discussed in the paper are the total mass, $M_{200}$ - enclosing 200 times the 
critical density, the corresponding radius, $R_{200}$, the cluster position defined as the location of the most bound particle, 
and the stellar mass defined as the stellar mass of the galaxy located in the most massive subhalo 
(i.e the central galaxy, or Bright Cluster Galaxy, BCG).

The main galaxy sample used in the paper is composed of the satellite galaxies of the cluster-like haloes defined above. 
The way \textsc{Subfind} stores the information of the galaxy population allows us to select them either using the default 
satellite definition or considering all galaxies - central and satellites - that are located within a given distance 
from the host halo centre. The existence of a splashback radius \citep[e.g][]{more2015,diemer2017,busch&white2017,baxter2017} shows that 
the virial radius does not represent a physical boundary of a cluster, so properties of galaxies can be affected by the environment 
beyond this limit as well as by variations of the density profile or infall material related to the filamentary environment within which 
clusters live. As shown in \citetalias{niemiec2019} using the Illustris simulation, tidal stripping can affect the subhaloes properties 
starting at $\sim 2 \times R_{200}$,  we therefore chose this value as our fiducial boundary for selecting satellite galaxies. In this 
paper, to take into account projection effects as they can affect observations,  we sometimes select galaxies within $2\times R_{200}$ 
\emph{projected}, keeping galaxies that are located at $\pm$ 5Mpc in the cluster line-of-sight. For instance at redshift $z=0.2$ this corresponds 
to a redshift uncertainty of approximately $\delta z = 0.001$.  
In Table~\ref{tab:sat_num} we report the total number of resolved satellites (i.e with $m_{\star} > 10^9 h^{-1}M_{\odot}$, and size $> 2h^{-1}\rm{kpc}$) at the present time $z=0$, according to these 
two definitions: using 3D and 2D selection.

We use different properties and mass definitions for satellite galaxies throughout this work, 
as stored in the \textsc{Subfind} catalogue, we consider:
\begin{itemize}
    \item $m_{\mathrm{sub}}$: the total mass of a subhalo, defined as the sum of the mass of all gravitationally bound particles;
    \item $m_{\mathrm{DM}}$: the mass of all the gravitationally bound dark matter particles of a subhalo;
    \item $m_{\star}$: the stellar mass of a satellite galaxy, defined as the mass of the star particles contained in twice the half-light radius of the galaxy;
    \item $m_{\star}^{\mathrm{corr}}$: resolution effects have shown to impact the stellar mass of galaxies in the TNG simulation \citep{weinberger2018, pillepich2018a}. At a given (sub)halo mass, galaxies have a lower stellar mass as measured in the less resolved but higher volume simulation run TNG300-1, compared to TNG100-1. Following the overall method described in \citet{pillepich2018b, engler2021} to correct for this effect, we use the low resolution version of TNG100-1, TNG100-2, and correct the stellar mass measured in TNG300 following Eq.~A1 from \citet{pillepich2018b}, as
    \begin{equation}
        m_{\star}^{\mathrm{corr,TNG300}} (m_{\mathrm{sub}}) = m_{\star}^{\mathrm{TNG300}} (m_{\mathrm{sub}}) \times \frac{<m_{\star}^{\mathrm{TNG100-1}} (m_{\mathrm{sub}})>}{<m_{\star}^{\mathrm{TNG100-2}} (m_{\mathrm{sub}})>},
    \end{equation}    
    where $<m_{\star}^{\mathrm{TNG100-1}} (m_{\mathrm{sub}})>$ and $<m_{\star}^{\mathrm{TNG100-2}} (m_{\mathrm{sub}})>$ are the mean stellar masses computed in subhalo mass bins in the TNG100-1 and TNG100-2 simulations respectively, TNG100-2 having the same volume as TNG100-1 but same resolution as TNG300.
    \item $m_{\mathrm{acc}}$: mass (subhalo or stellar) at the time of accretion, defined as the time when the subhalo first enters within $2\times R_{200}$ from the host center;
    \item sSFR: instantaneous star formation rate (SFR), which is the sum of the SFR in all the galaxy gas cells, divided by the stellar mass. To be consistent with the chosen definition of $m_{\star}$, the sum is taken over all the cells included within twice the half-light radius. As the SFR is not resolved for all galaxies (and then given as zero), for the non resolved galaxies we draw random SFR values between $10^{-4}$ and $10^{-5}M_{\odot}/yr$ \citep[following for instance][]{donnari2021a};
    \item galaxy size: we use the half-mass radius computed for the star particles.
\end{itemize}

As the cluster environment can affect the star formation rate of satellite galaxies, and quenched galaxies represent the main population in clusters, we study these populations separately in parts of our analysis. We therefore split our satellite galaxies into an active and a passive sample at $z=0$, with the limit $\mathrm{sSFR}_{\mathrm{lim}} =  10^{-11}yr^{-1}$, which is a threshold value typically adopted in the literature for low redshift studies \citep[e.g][]{wetzel2013, bahe2014}. We note that the star formation rate of galaxies is also affected by the simulation resolution, as demonstrated in \citet{pillepich2018a}. However, as both the SFRs and stellar masses are affected, the resulting sSFRs are only marginally impacted, which we verify by comparing sSFRs distributions for galaxies from the TNG100-1 and TNG100-2 simulation runs. We select active galaxies according to their sSFRs as measured in TNG100-2, and compare with the selection when correcting the sSFRs to match that of galaxies in TNG100-1, and find that only $\sim 1\%$ of galaxies are misclassified, which does not affect our results.
In Table~\ref{tab:sat_num}, we report 
the number of galaxies in each sample.

\begin{table}
    \centering
    \begin{tabular}{|c|c|c|c|c|}

        \hline
        \multicolumn{5}{|c|}{Within $2 \times R_{200}^{\mathrm{3D}}$}              \\
        \hline
        $m_{\star}$ bin & All   & Active    & Passive   & \% passive\\
        \hline
        All            & 14820  & 1704      & 13116     & 89        \\
        $[9-9.5]$      & 5676   & 703       & 4973      & 88        \\
        $[9.5-10]$     & 4380   & 717       & 3663      & 85        \\
        $[10-10.5]$    & 3626   & 262       & 3364      & 93        \\
        $[10.5-11]$    & 959    & 13        & 946       & 99        \\
        $[11-11.5]$    & 167    & 9         & 158       & 95        \\
        $[11.5-12]$    & 11     & 0         & 11        & 100       \\
        \hline 
        
        \hline
        \multicolumn{5}{|c|}{Within $2 \times R_{200}^{\mathrm{proj}}$}              \\
        \hline
        $m_{\star}$ bin & All   & Active    & Passive   & \% passive\\
        \hline
        All            & 17228  & 2882      & 14346     & 83        \\
        $[9-9.5]$      & 6613   & 1250      & 5363      & 81        \\
        $[9.5-10]$     & 5140   & 1204      & 3936      & 77        \\
        $[10-10.5]$    & 4170   & 400       & 3770      & 90        \\
        $[10.5-11]$    & 1085   & 16        & 1069      & 99        \\
        $[11-11.5]$    & 204    & 12        & 192       & 94        \\
        $[11.5-12]$    & 15     & 0         & 15        & 100       \\
        \hline 
    \end{tabular}
    \caption{Number of resolved satellite galaxies, i.e. galaxies with i.e with $m_{\star} > 10^9 h^{-1}M_{\odot}$, and size $> 2h^{-1}\rm{kpc}$ located within $2 \times R_{200}$ from their host centre, with contamination from projection effects (\textit{bottom}) and without (\textit{top}). The stellar mass ranges are expressed in $\log (m_{\star} h^{-1}M_{\odot})$. }
    \label{tab:sat_num}
\end{table}

\section{Obtaining $m_{\mathrm{sub}}$ from $m_{\star}$}
\label{sec:shmr_obs}
\subsection{Full SHMR}
\label{sec:full_shmr_obs}

We first look at the SsHMR from an \emph{observational point of view}, i.e. we measure the total mass of subhaloes 
$m_{\mathrm{sub}}$ as a function of the satellite galaxy stellar masses $m_{\star}$. 
In this section, we present fitting functions for the median subhalo mass as a function of galaxy stellar mass, 
and for the scatter around this median relation. Given these, and defining the median subhalo mass 
as $m_{\mathrm{sub}}^{\mathrm{med}}(m_{\star})$ and scatter as $\delta m_{\mathrm{sub}}$, 
we can obtain the subhalo masses for an (observed) cluster given the stellar masses as:
\begin{equation}
    m_{\mathrm{sub}}(m_{\star}) = m_{\mathrm{sub}}^{\mathrm{med}}(m_{\star}) + \delta m_{\mathrm{sub}},
\end{equation}
where $\delta m_{\mathrm{sub}}$ follows a probability distribution $p_{\mathrm{Msub}}$ that we intent to constrain.

\begin{figure} 
  \begin{center}
    \includegraphics{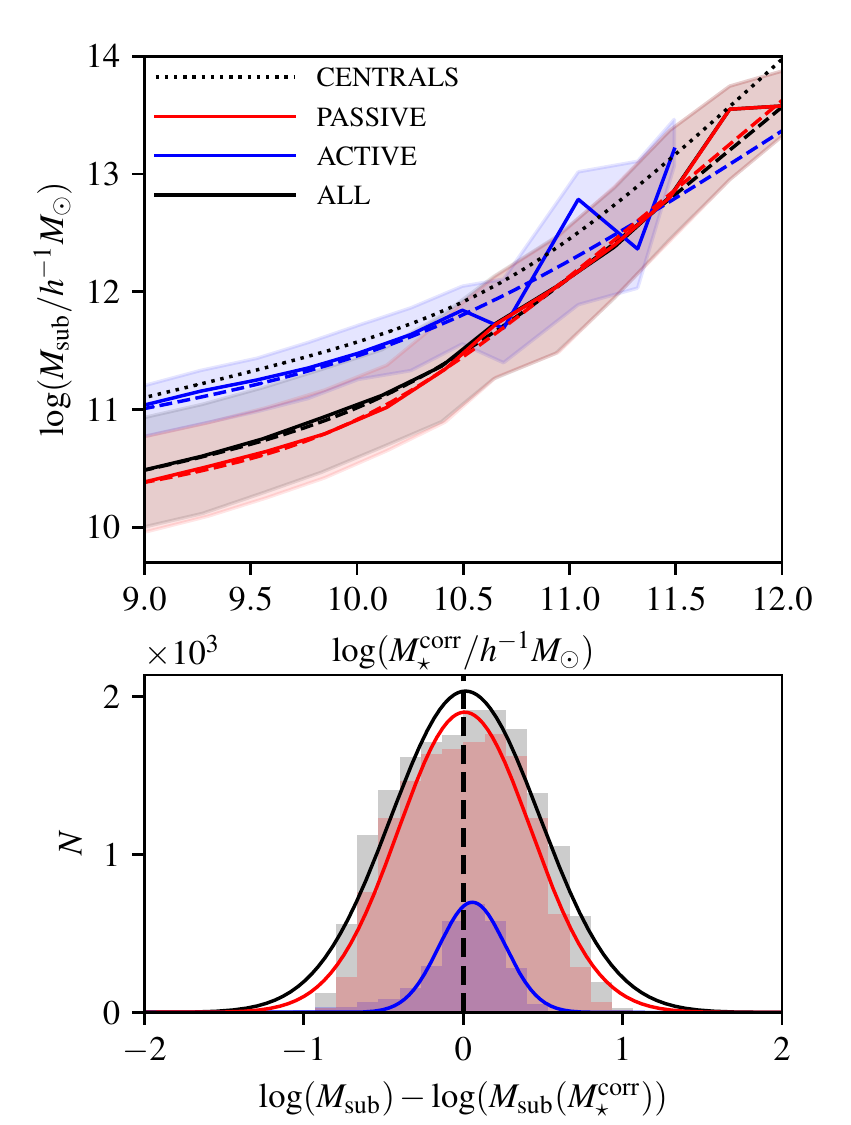}
    \caption{\textit{Top panel:} Stellar-to-Halo Mass Relation for the satellite galaxies within $2\times R_{200}$ (projected) of the cluster centre, for all galaxies (black), and active (blue) and passive (red) separately. The solid line shows the median relation, and the $16^{th}-84^{th}$ percentiles are shown as the shaded regions. The dashed lines show the best fit relations. As a comparison, the dotted line shows the fit for all the central galaxies of the simulation.  \textit{Bottom panel: }residual distribution for the full, active and passive samples, in black, blue and red respectively.}
		\label{fig:shmr_residuals}
  \end{center}
\end{figure}

We remind the reader that hereafter $m_{\star} \equiv m_{\star}^{\mathrm{corr}}$, as described in Sect.~\ref{sec:sample}, and 
that we select all galaxies  located within 
$2 \times R_{200}^{2D}$ (\emph{projected}) from the cluster centre, and $\pm 5$\,Mpc along the line-of-sight. We note that for each cluster we only account for one 2D projection.
We keep only well resolved galaxies, i.e. with $m_{\star} > 10^9 h^{-1}M_{\odot}$, and size $> 2h^{-1}\rm{kpc}$. 
The measured SsHMRs for satellite galaxies from TNG300 are shown on the top panel of Fig.~\ref{fig:shmr_residuals} 
for the whole sample in black, for passive galaxies in red and star-forming ones in blue (as defined in Sect.~\ref{sec:sample}). The solid lines and shaded regions represent the median relations, and enclose the $16^{th}-84^{th}$ percentile regions. 

We model the relation considering a double power law function to the SsHMR, as done by \citet{moster2013}: 
\begin{equation}
    m_{\mathrm{sub}} (m_{\star}) = 2N\left[\left(\frac{m_{\star}}{M_1}\right)^{-\beta} + \left(\frac{m_{\star}}{M_1}\right)^{\gamma}\right] m_{\star}.
    \label{equ:SsHMR}
\end{equation}
We fit the function for the whole satellite sample, then for the active and passive galaxy sample separately, with a Markov Chain Monte Carlo (MCMC) method using \textsc{emcee} \citep{foreman-mackey2013}, a Python implementation of an affine invariant MCMC ensemble sampler. 
The best fit parameters and 68\% intervals of confidence are given in Table~\ref{tab:shmr}, and the corresponding relations are 
displayed as dashed lines in the top panel of Fig.~\ref{fig:shmr_residuals}. 
For comparison, using Eq.~\ref{equ:SsHMR} we model all central galaxies of the simulation with $M_{\star} > 10^9h^{-1}M_{\odot}$, and show the best fit relation using dotted curve in the top panel of Fig.~\ref{fig:shmr_residuals}. The best fit parameters for Eq.~\ref{equ:SsHMR} for central galaxies are $N^{\mathrm{cent}} = 6.32^{+0.06}_{-0.06}$; $\log M_1^{\mathrm{cent}} = 10.73^{+0.02}_{-0.02}$; $\beta^{\mathrm{cent}} = 0.58^{+0.01}_{-0.01}$; $\gamma^{\mathrm{cent}} = 0.68^{+0.03}_{-0.03}$\,.

As the scatter around the median/best fit relation does not vary significantly with stellar mass for the three satellite populations, we only parametrize the total scatter around the best fit relation. We show the residual distributions, $ \log(m_{\mathrm{sub}}) - \log(m_{\mathrm{sub}}(m_{\star}))$, in the bottom panel of Fig.~\ref{fig:shmr_residuals}, and model them with a normal distribution. The probability distribution parametrizing the scatter around the best fit relation, $p_{\mathrm{Msub}}$, is then given as a normal distribution with mean $m$ and standard deviation $\sigma$. The fitted values for the scatter parameters are given in Table~\ref{tab:shmr}. We remind that we only include the scatter for galaxies with $m_{\star} > 10^9 h^{-1}M_{\odot}$ and size $> 2h^{-1}\rm{kpc}$, considered as the "resolved" sample. We note that for the full and passive galaxy samples, the residual distribution may not seem to follow a normal distribution, but this appears to be mostly due to the discreetness of the data, as the simulation has a finite resolution. We compute the residuals using galaxies from the more resolved TNG100 run, and use a KS test to verify that they do follow a normal distribution: we find a KS statistic of $\sim 0.02$ for both samples, with a p-value of $\sim 0.6$ and $\sim 0.9$ for the full and passive sample respectively.

\begin{figure} 
  \begin{center}
    \includegraphics{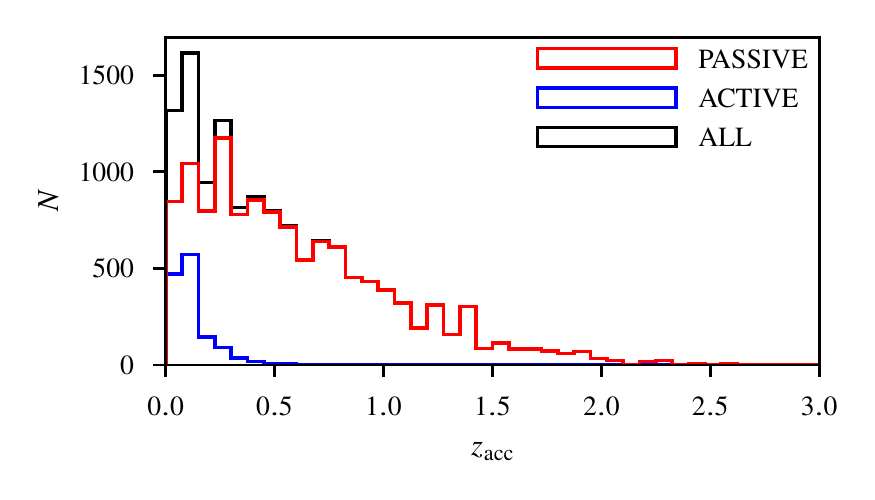}
    \caption{Accretion redshift distribution for all (black), active (blue) and passive (red) satellite galaxies. }
		\label{fig:zacc_hist}
  \end{center}
\end{figure}

The population of active galaxies in clusters follows a distinct SsHMR, as they are on a different evolutionary stage in the cluster compared to the passive sample. Indeed, on average these galaxies have  spent less time in the cluster and have therefore not been quenched by this dense environment yet: we show on Fig.~\ref{fig:zacc_hist} the distribution of redshift of accretion at $2\times R_{200}$ for all (black), active (blue) and passive (red) galaxies with $m_{\star} > 10^9 h^{-1}M_{\odot}$.
The average accretion redshift  for active galaxies is $<z_{\mathrm{acc}}> \sim 0.1$, while for the passive sample it is $<z_{\mathrm{acc}}> \sim 0.7$. 
Due to this, tidal stripping has not yet affected their subhaloes as much as their quenched counterparts: at a given stellar mass, active galaxies have a higher subhalo mass than passive ones, and follow a very similar SHMR to central galaxies. In particular, this is interesting because the relation is reversed for central/field galaxies \citep[e.g][]{mandelbaum2016}, meaning that at given stellar mass, red galaxies tend to have a more massive halo than blue ones, which illustrates that different processes govern the galaxy-halo connection for central and satellite galaxies. 
We note that some galaxies can also be already quenched at their time of accretion into the cluster, due to internal processes, or to preprocessing in other structures (see Sect.~\ref{sec:preprocessing}.)
 
To quantify the shift in SHMR between the central and satellite galaxy samples, we defined in \citetalias{niemiec2019} the stripping factor, $\tau_{\mathrm{strip}}$, as:
\begin{equation}
     \tau_{\mathrm{strip}}(m_{\star}) = 1 - \frac{\bar{m_{\mathrm{sub}}}(m_{\star})}{\bar{M_{\mathrm{h}}}(M_{\star})}.
     \label{equ:tau_strip}
\end{equation}
We measure the stripping factor in stellar mass bins, and $\bar{m_{\mathrm{sub}}}(m_{\star})$ ($\bar{M_{\mathrm{h}}}(M_{\star})$) represent the median subhalo (halo) mass for a bin of satellite (central) galaxies with median stellar mass $m_{\star}$ ($M_{\star}$). This definition comes from the fact that the main interaction affecting galaxies in clusters (as compared to central or field galaxies) is partial stripping of their dark matter by tidal forces of hosts. Defined as it is, the stripping factor does not aim at accounting for all physical processes that impact satellite galaxies since their infall; it is rather a simple observable, that can be used to quantify the difference in evolution for central and satellite galaxies, and compare different galaxy samples. Passive galaxies are more stripped than active galaxies, which is reflected in their stripping factor: $<\tau_{\mathrm{strip}}^{\mathrm{active}}> = 0.50 \pm 0.20$ while $<\tau_{\mathrm{strip}}^{\mathrm{passive}}> = 0.78 \pm 0.06$. For the total galaxy population, the shift in the SHMR is driven by the passive galaxy population (<$\tau_{\mathrm{strip}}^{\mathrm{all}}> = 0.78 \pm 0.06$). We present the stripping factor as a function of the stellar mass for the full (black), active (blue) and passive (red) satellite galaxies in the left panel of Fig.~\ref{fig:tau_strip}.
 
We now want to use some additional observable parameters  to parametrize the SsHMR in order to reduce and regulate the measured scatter.
As the SsHMR followed by the active galaxy sample has a quite low scatter ($\sigma_{\mathrm{active}} = 0.21$ vs $\sigma_{\mathrm{passive}} = 0.45$), we will look at the dependence on additional parameters only for the passive galaxy SsHMR.

\begin{table*}
    \centering
\subfloat[]{\begin{tabular}{|c|c|c c c c c c|c c|c|c|}
\multicolumn{11}{c}{$m_{\mathrm{sub}} = f(m_{\star})$} \\
\hline
                    & $z$   & $N$   & $\log M_1$    & $\beta$   & $\gamma$   & $a$  & $b$   & $m$ & $\sigma$  & $N_{\mathrm{sat}}$ & \% passive    \\
\hline
\multirow{3}{*}{All}& $0$   &  $2.38\,^{+0.11}_{-0.09}$ & $10.22\,^{+0.11}_{-0.12}$ & $0.65\,^{+0.05}_{-0.04}$  & $0.50\,^{+0.10}_{-0.10}$  & -- & -- & 0.01 & 0.47 & 17228 & 83 \\
                    & $0.24$&  $2.29\,^{+0.09}_{-0.08}$ & $10.73\,^{+0.09}_{-0.10}$ & $0.47\,^{+0.03}_{-0.03}$  & $0.83\,^{+0.17}_{-0.16}$  & -- & -- & -0.02 & 0.46 & 11259 & 76  \\
                    & $0.5$ &  $2.20\,^{+0.11}_{-0.10}$ & $10.60\,^{+0.13}_{-0.17}$ & $0.52\,^{+0.05}_{-0.04}$  & $0.66\,^{+0.20}_{-0.19}$  & -- & -- & -0.01 & 0.46 & 6965 & 71 \\
\hline
\multirow{3}{*}{Active} & $0$ & $5.49\,^{+3.68}_{-1.54}$& $10.34\,^{+0.52}_{-0.60}$ & $0.70\,^{+0.15}_{-0.11}$ & $0.19\,^{+0.59}_{-0.29}$ & -- & -- & 0.06 & 0.20 & 2882 & -- \\
                    & $0.24$& $4.06\,^{+0.67}_{-0.51}$  & $10.74\,^{+0.18}_{-0.30}$ & $0.59\,^{0.08}_{-0.05}$   & $0.60\,^{+0.46}_{-0.36}$ & -- & -- & 0.06 & 0.28 & 2666 & -- \\
                    & $0.5$ & $3.60\,^{+0.66}_{-0.49}$  & $10.72\,^{+0.19}_{-0.34}$ & $0.59\,^{+0.08}_{-0.06}$  & $0.56\,^{+0.50}_{-0.39}$ & -- & -- & 0.05 & 0.29 & 2050 & -- \\
\hline
\multirow{3}{*}{Passive} & $0$ & $2.02\,^{+0.09}_{-0.07}$& $10.04\,^{+0.09}_{-0.10}$ & $0.72\,^{+0.05}_{-0.05}$ & $0.52\,^{+0.08}_{-0.08}$ & -- & -- & 0.01 & 0.42 & 14346 & -- \\
                    & $0.24$& $1.86\,^{+0.08}_{-0.07}$  & $10.25\,^{+0.11}_{-0.13}$ & $0.61\,^{+0.05}_{-0.04}$  & $0.58\,^{+0.12}_{-0.11}$ & -- & -- & -0.02 & 0.39 & 8593 & --  \\
                    & $0.5$ & $1.73\,^{+0.14}_{-0.10}$ & $10.11\,^{+0.15}_{-0.17}$  & $0.69\,^{+0.08}_{-0.06}$  & $0.51\,^{+0.14}_{-0.12}$  & -- & -- & -0.02 & 0.39 & 4915 & -- \\
\hline
\multirow{3}{*}{Passive with $x_{\mathrm{sat}}^{\mathrm{2D}}$} & $0$ & -- & $10.08\,^{+0.11}_{-0.11}$ & $0.66\,^{0.05}_{-0.05}$ & $0.51\,^{+0.09}_{-0.08}$ & $2.19\,^{+0.12}_{-0.11}$ & $0.77\,^{+0.05}_{0.05}$ & 0.01 & 0.30& 14346 & -- \\
                    & $0.24$& -- & $10.32^{+0.12}_{-0.14}$  & $0.56^{+0.05}_{-0.04}$ & $0.60\,^{+0.13}_{-0.12}$ & $2.12\,^{+0.14}_{-0.13}$  & $0.75\,^{+0.06}_{-0.05}$ & -0.03 &  0.31 &  8593 & -- \\
                    & $0.5$ & -- & $10.23^{+0.17}_{-0.20}$  & $0.60^{+0.07}_{-0.06}$ & $0.53\,^{+0.16}_{-0.15}$ & $2.02\,^{+0.20}_{-0.18}$  & $0.76^{+0.08}_{-0.07}$ & -0.02 &  0.32 & 4915 & -- \\
\hline
\end{tabular}}

\subfloat[]{\begin{tabular}{|c|c|c c c c c c c|c c|c|}
\multicolumn{11}{c}{$m_{\star} = f(m_{\mathrm{sub}})$} \\
\hline
    & $z$   & $N'$  & $M1'$ & $\beta'$  & $\gamma'$ & $A$   & $c$   & $d$   & $m$ & $\sigma$ & $N_{\mathrm{sat}}$ \\ 
\hline
\multirow{3}{*}{All} & 0 & $0.056\,^{+0.005}_{-0.005}$  & $11.14\,^{+0.09}_{-0.09}$    & $1.24\,^{+0.13}_{-0.11}$     & $0.35\,^{+0.07}_{-0.07}$     & -- & -- & -- & 0.01 & 0.58 & 19341 \\
& 0.24  & $0.042_{-0.004}^{0.005}$  & $11.05_{-0.07}^{0.08}$    & $1.56_{-0.14}^{0.15}$ & $0.24_{-0.07}^{0.07}$   & --    & --    & --    & 0.01 & 0.65 & 12663 \\
& 0.5   & $0.057_{-0.006}^{0.006}$  & $11.30_{-0.10}^{0.10}$    & $1.27_{-0.11}^{0.13}$  & $0.43_{-0.10}^{0.11}$   & --    & --    & --    & -0.01 & 0.66    & 7775 \\
\hline
\multirow{3}{*}{With $x_{\mathrm{sat}}^{\mathrm{3D}}$}    & 0 & $0.048\,^{+0.004}_{-0.004}$ & $11.20\,^{+0.08}_{-0.07}$   & --    & $0.32\,^{+0.07}_{-0.07}$ & $-1.03\,^{+0.04}_{-0.04}$ & $0.47\,^{+0.09}_{-0.08}$ & $0.76\,^{+0.09}_{-0.08}$ & 0.04 & 0.45 & 19341 \\
& 0.24  & $0.046_{-0.004}^{0.004}$  & $11.29_{-0.08}^{0.08}$    & --    & $0.35_{-0.07}^{0.08}$ & $-1.09_{-0.05}^{0.05}$  & $0.54_{-0.07}^{0.08}$ & $0.63_{-0.08}^{0.08}$ & -0.01 & 0.50  & 12663  \\
& 0.5   & $0.052_{-0.005}^{0.005}$  & $11.44_{-0.09}^{0.10}$    & --    & $0.45_{-0.10}^{0.11}$   & $-1.02_{-0.07}^{0.07}$    & $0.51_{-0.07}^{0.08}$ & $0.58_{-0.08}^{0.08}$   & -0.01 & 0.48  & 7775    \\
\hline
\multirow{1}{*}{With $x_{\mathrm{min}}$}    & 0 & $0.024^{+0.003}_{-0.002}$ &  $11.08^{+0.08}_{-0.07}$ & -- & $0.12^{+0.06}_{-0.06}$   & $-1.06^{+0.03}_{-0.03}$   & $0.55\,^{+0.13}_{-0.11}$    & $0.94\,^{+0.08}_{-0.07}$  & 0.01 & 0.34  & 19341 \\
\hline

     \end{tabular}}
    \caption{Best fit parameters for the SsHMR median relations and scatter. The top table gives the parameters for $m_{\mathrm{sub}} = f(m_{\star})$ as described in Sect.~\ref{sec:shmr_obs}, where $N_{\mathrm{sat}}$ is the number of satellite galaxies used for each fit, that is with $m_{\star} > 10^9h^{-1}M_{\odot}$ and $R_{\mathrm{sat}}^{\mathrm{2D}} < 2\times R_{200}$. The fraction of passive satellite galaxies within $2\times R_{200}$ projected is 83, 76 and 70\% at redshift $z =0$, 0.24 and 0.5 respectively. The bottom table gives the parameters for $m_{\star} = f(m_{\mathrm{sub}})$ as described in Sect.~\ref{sec:shmr_simu}: from top to bottom, best fit parameters for the SsHMR $m_{\star} = f(m_{\mathrm{sub}})$ from Equation \ref{equ:shmr_simu}, $m_{\star} = f(m_{\mathrm{sub}}, x_{\mathrm{sat}}^{\mathrm{3D}})$ from Eq.~\ref{equ:shmr_simu_Rsat} and $m_{\star} = f(m_{\mathrm{sub}}, x_{\mathrm{min}})$, and width of the total scatter around these relations.}
    \label{tab:shmr}
\end{table*}

    \subsection{Radial dependence of the SHMR}
    \label{sec:sshmr_rsat}
    
To understand and reduce the scatter in the SsHMR, we first look at its evolution with the distance to the cluster centre. As galaxies tend to fall with time towards the centre of their host due to dynamical friction \citep[see for instance][]{nipoti2017}, it can be expected that galaxies located closer to the core will have different properties than the ones located in the outskirts, as they have been subject to  more interactions within their surrounding dense environment \citep[e.g][]{rhee2017, gu2020}.

As mentioned in the previous section, we only focus on the passive galaxy sample.   
To mimic what can be obtained with observations, we take into account projection effects: cluster galaxies are selected within $2\times R_{200}$ \emph{projected} (see Sect.\ref{sec:sample}), and the variations of the SsHMR are examined with respect to the projected 2D distance to the cluster center. To reduce differences due to the range in cluster masses, we normalize the cluster-centric distance by the virial radius of the cluster at redshift zero, thus:
\begin{equation}
    x_{\mathrm{sat}}^{\mathrm{2D}} \equiv R_{\mathrm{sat}}^{\mathrm{2D}}/R_{200}.
\end{equation}

\begin{figure} 
  \begin{center}
    \includegraphics{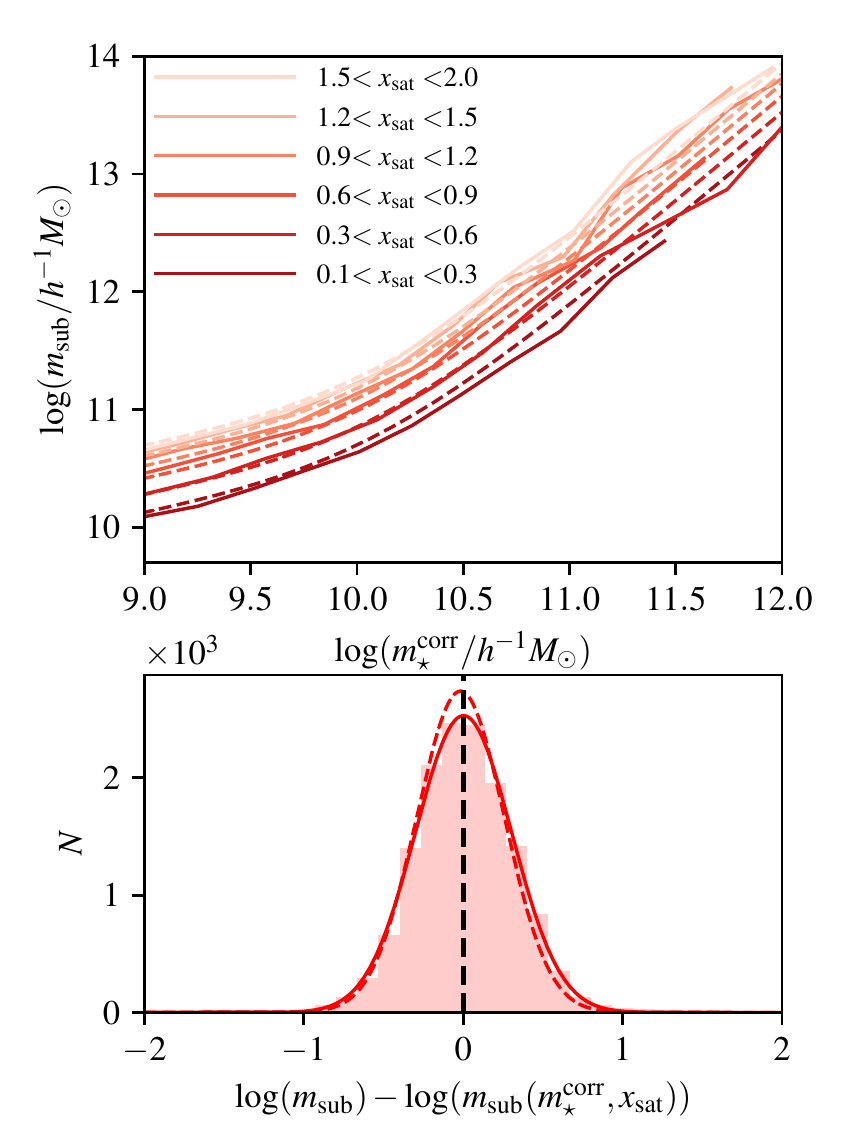}
    \caption{\textit{Top panel:} stellar-to-halo mass relation for the passive satellite galaxies, split in bins of projected distance to the cluster centre. The solid line shows the median relation, and the 16$^{th}$-84$^{th}$ percentiles are shown as the shaded regions. The dotted line shows the best fit relations. \textit{Bottom panel: } distribution of residuals with respect to the best fit function. We fit a Gaussian to the distribution, shown as the solid line. Then the dashed line shows the shape of the residual distribution considering the 3D distance to the cluster distance in equation \ref{equ:SsHMR_rsat} instead of the projected distance. }
		\label{fig:shmr_rsatbins}
  \end{center}
\end{figure}

We then add a simple linear dependence on $x_{\mathrm{sat}}^{\mathrm{2D}}$ to equation \ref{equ:SsHMR}, which gives:
\begin{equation}
    m_{\mathrm{sub}} (m_{\star}, x_{\mathrm{sat}}^{\mathrm{2D}}) = 2\left[\left(\frac{m_{\star}}{M_1}\right)^{-\beta} + \left(\frac{m_{\star}}{M_1}\right)^{\gamma}\right] m_{\star} 
    \times (a x_{\mathrm{sat}}^{\mathrm{2D}} + b), 
    \label{equ:SsHMR_rsat}
\end{equation}
and fit the new parameters $M_1$, $\beta$, $\gamma$, $a$ and $b$ with the \textsc{emcee} MCMC method. The best fit values, along with the 16th-84th confidence intervals are given in Table~\ref{tab:shmr}.
We show the measured and best fit SsHMR in bins of $x_{\mathrm{sat}}^{\mathrm{2D}}$ in the top panel of Fig.~\ref{fig:shmr_rsatbins}. The SsHMR shows a clear evolution with $x_{\mathrm{sat}}^{\mathrm{2D}}$: at a given stellar mass, galaxies closer to the cluster core live on average in a less massive subhalo than galaxies that live in the outskirts. We will examine the processes that lead to this effect in Sect.~\ref{sec:time_evol}.
We measure the stripping factor as defined in equation ~\ref{equ:tau_strip} in three bins of projected cluster-centric distance normalized by $R_{200}$, $x_1 = [0.1; 0.5]$, $x_2 = [0.5; 1]$ and $x_3 = [1; 2]$. The evolution of the stripping factor as a function of the stellar mass for the three bins in shown in the left panel of Fig.~\ref{fig:tau_strip} with dashed lines.
As expected, galaxies closer to the cluster centre have higher stripping factors: $\tau_{\mathrm{strip}}^{\mathrm{x1}} = 0.88 \pm 0.04$, $\tau_{\mathrm{strip}}^{\mathrm{x2}} = 0.74 \pm 0.06$ and $\tau_{\mathrm{strip}}^{\mathrm{x3}} = 0.56 \pm 0.13$.

We also measure the residuals with respect to the best fit relation for the full passive galaxy sample, $\log m_{\mathrm{sub}} - \log m_{\mathrm{sub}}(m_{\star}^{\mathrm{corr}}, x_{\mathrm{sat}}^{\mathrm{2D}})$, and fit a Gaussian to the distribution. This is shown as a solid line in the bottom panel of Fig~\ref{fig:shmr_rsatbins}. Parametrizing  the SsHMR with $x_{\mathrm{sat}}^{\mathrm{2D}}$ allows to reduce the scatter: it goes from $\sigma = 0.42$ for the full SsHMR for the passive galaxy sample, to $\sigma = 0.30$ for the radially-dependant SsHMR.

As described above, we use the projected radial distance in order to mimic what is possible in observations, but this can add scatter in the radial SsHMR. To test the impact of this projection effect, we fit again equation \ref{equ:SsHMR_rsat} to the simulated galaxies, but using the 3D distance, $x_{\mathrm{sat}}^{\mathrm{3D}}$, instead of the projected $x_{\mathrm{sat}}^{\mathrm{2D}}$. We find $M_1^{\mathrm{3D}} = 10.09_{-0.14}^{+0.12}$, $\beta^{\mathrm{3D}} = 0.65_{-0.05}^{+0.06}$, $\gamma^{\mathrm{3D}} = 0.47_{-0.09}^{+0.10}$, $a^{\mathrm{3D}} = 2.19_{-0.12}^{+0.14}$ and $b^{\mathrm{3D}} = 0.46_{-0.04}^{+0.05}$. The shape of the residual distribution is shown with dashed line in the bottom panel of Fig.~\ref{fig:shmr_rsatbins}, and the scatter is indeed slightly decreased compared to the 2D fit, from $\sigma_{\mathrm{2D}} = 0.30$ to $\sigma_{\mathrm{3D}} = 0.27$. However, this shows that projection effects are not the dominant cause of the remaining scatter in the radial SsHMR, but rather that there is an intrinsic variation in the SsHMR at each distance to the cluster centre.

    \subsection{Influence of other observable parameters}
    \label{sec:shmr_obs_params}
    
To better understand the remaining scatter in the radial SsHMR, we examine the impact that other observable parameters have on this relation.
To achieve this, we look at the correlations between residuals of the radial SsHMR, $\Delta (\log m_{\mathrm{sub}}) = \log m_{\mathrm{sub}} - \log m_{\mathrm{sub}}(m_{\star}^{\mathrm{corr}}, x_{\mathrm{sat}}^{\mathrm{2D}})$, and observable parameters that could impact the evolution of subhaloes/satellite galaxies in a cluster. We have already included the main parameters related to subhalo evolution, the radial distance to cluster centre and the specific star formation rate (active vs passive). Another parameter that could have an impact is the size of the satellite galaxies: at a given stellar mass, more extended galaxies could be more affected by tidal stripping than more compact ones. At the same time, galaxy evolution could depend on the properties of the host cluster. The cluster properties that we consider, and that could have an impact on the satellite galaxies, are the total cluster mass, $M_{200}$, the mass of the BCG normalized by the cluster mass, $M_{\star, \mathrm{BCG}}/M_{200}$, the concentration $c_{200}$, and the time of cluster formation, $z_{\mathrm{form}}$, defined as the redshift at which the cluster had assembled half of his redshift zero mass \citep[concentrations and formation redshifts obtained from][]{anbajagane2022}. 

\begin{figure*} 
  \begin{center}
    \includegraphics{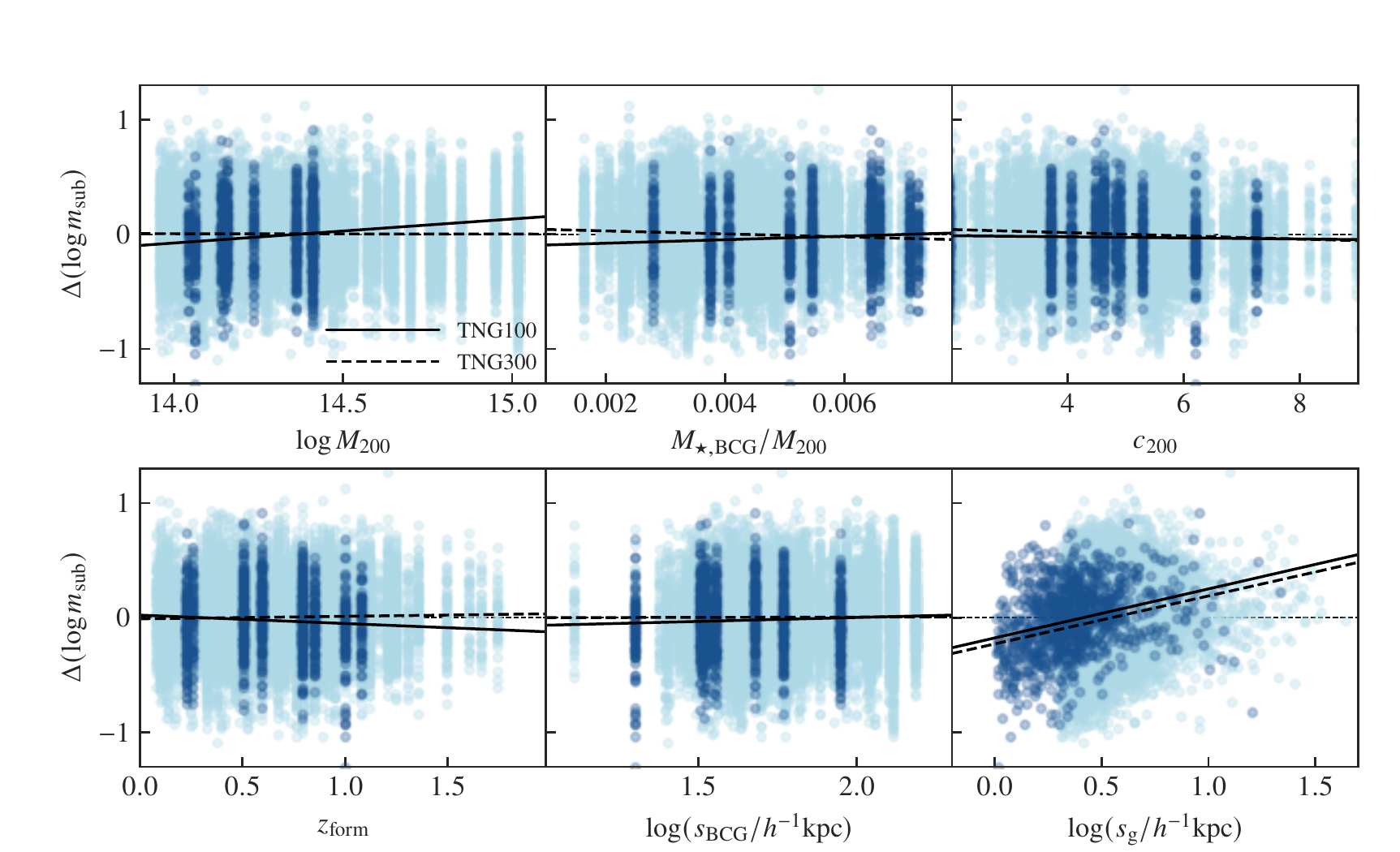}
    \caption{Residuals $\Delta(\log m_{\mathrm{sub}}) = \log m_{\mathrm{sub}} - \log m_{\mathrm{sub}}(m_{\star}, x_{\mathrm{sat}}^{\mathrm{2D}})$ as a function of (from left to right and top to bottom) the host mass $M_{200}$, the ratio between the mass of the BCG and of the host, the host concentration, host formation redshift, size of the host BCG and galaxy size. We show the full distribution of residuals as dots, and a linear fit as a line, for satellite galaxies from the TNG300-1 simulation (light blue and dashed line), and TNG100-1 simulation (dark blue and solid line). All masses are expressed in $h^{-1}M_{\odot}$.}
		\label{fig:all_residuals}
  \end{center}
\end{figure*}

\begin{figure*} 
  \begin{center}
    \includegraphics{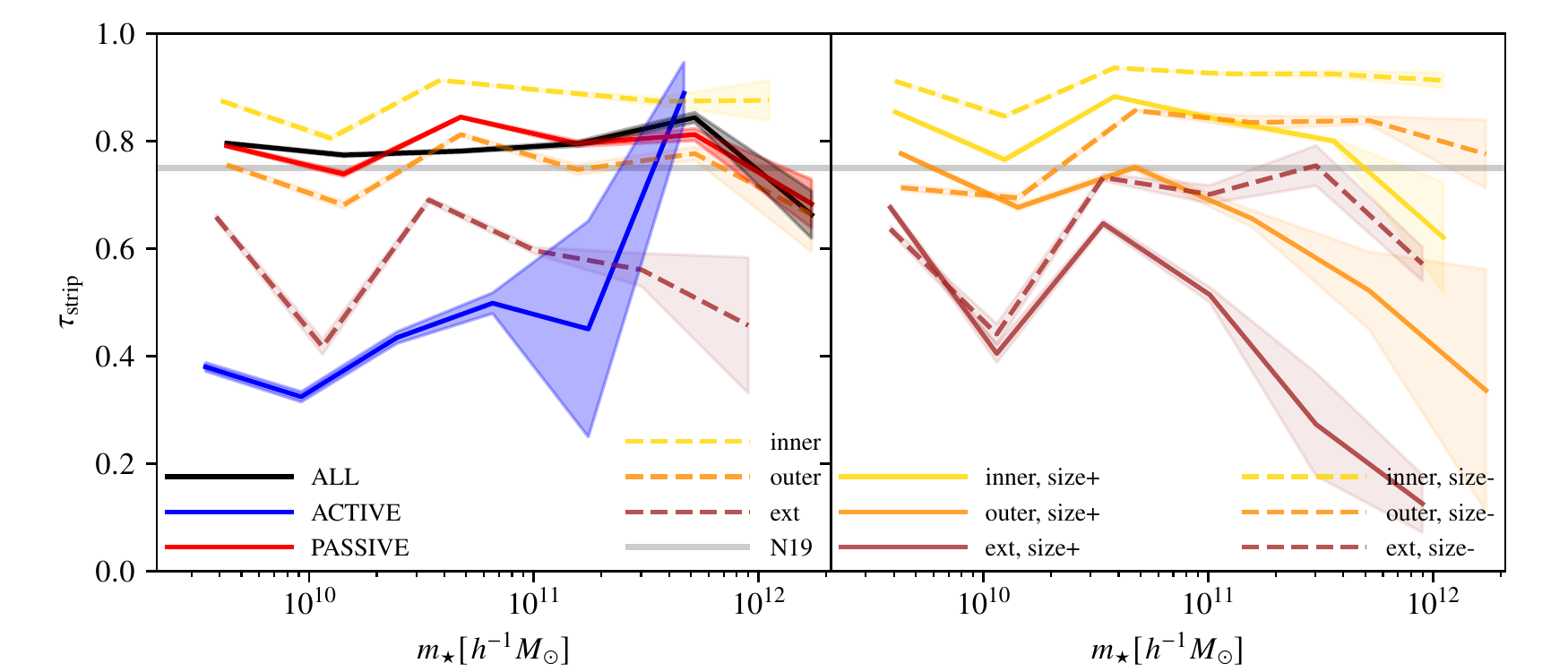}
    \caption{Stripping factor as a function of the stellar mass for different galaxy selections. \textit{Left panel}: for all satellite galaxies from TNG300 with $m_{\star} > 10^9 h^{-1}M_{\odot}$, and located within $2\times R_{200}$ projected from he cluster centre, and $\pm 5$\,Mpc along the line of sight, shown as a black curve. The stripping factor is also shown for the active (blue) and passive (red) satellite populations. The passive galaxies are further split into bins in projected distance to cluster core, where the inner ($0.1 < R_{\mathrm{sat}}^{\mathrm{2D}}/R_{200}$ < 0.5) sample is shown in yellow, the outer ($0.5 < R_{\mathrm{sat}}^{\mathrm{2D}}/R_{200}$ < 1) in orange and the external ($1 < R_{\mathrm{sat}}^{\mathrm{2D}}/R_{200}$ < 2) in brown. \textit{Right panel:} stripping factor for the same bins in projected distance to the cluster centre, but further split according to the galaxy sizes at a given stellar mass. More compact galaxies (i.e smaller in size) are shown as dashed lines, and more extended galaxies with solid lines. }
		\label{fig:tau_strip}
  \end{center}
\end{figure*}

Figure \ref{fig:all_residuals} shows the correlation between residuals $\Delta (\log m_{\mathrm{sub}})$ and the 6 parameters described above. Light blue points are the values for individual galaxies, and the dashed line is a linear fit to the data points. There is no significant dependence of the radial SsHMR on any of the galaxy cluster properties. Possibly, if there are some variations, they can be absorbed by the normalisation of the cluster-centric distance by the virial radius of the clusters $R_{200}$.
The parameter that seems to correlate with the radial SsHMR residuals is the size of satellite galaxies. It is important to note that this size dependence is a residual after taking into account the stellar mass dependence, which means that it is more a "compactness" dependence \citep[see][for the impact of galaxy compactness on halo mass in the case of central galaxies]{huang2020}: at a given stellar mass, a larger galaxy will follow a different SsHMR than a smaller one.  To check how  this galaxy compactness impacts the SsHMR, we measure again the stripping factor as defined in Equation \ref{equ:tau_strip} in three $x_{\mathrm{sat}}^{\mathrm{2D}}$ bins, but this time, within each stellar mass and $x_{\mathrm{sat}}^{\mathrm{2D}}$ bin, we subdivide galaxies into two samples, according to their size (compared to the median size over the $[m_{\star}, x_{\mathrm{sat}}^{\mathrm{2D}}]$ bin). For each stellar mass and cluster-centric distance, smaller galaxies have had a significantly higher stripping factor than larger ones. The stripping factor as a function of stellar mass for these galaxy samples is shown in the right panel of Fig.~\ref{fig:tau_strip}, with the relation for more compact galaxies shown in dashed lines, and for more extended ones shown as solid lines. This can be a direct result of the outer-in stripping process (i.e galaxies that end up more compact are the ones that have been stripped more), or can come from a different evolution of galaxies that are more or less compact prior to their accretion into the cluster. We investigate this further in Sect.~\ref{sec:time_evol}. 

To verify if there is no effect of the simulation resolution on these correlations, we compute the same residuals but for the satellite galaxies from TNG100, in which a dark matter particle is 8 times less massive than in TNG300, and gravitational softening length is about half compared to TNG300. These residuals are shown for galaxies with $m_{\star} > 10^9 h^{-1}M_{\odot}$ and size $> 1h^{-1}\rm{kpc}$,  as dark blue dots in Fig.~\ref{fig:all_residuals}, with a linear fit shown as solid line. There is not significant bias in the residual distribution computed for the satellite galaxies from TNG100.
Additionally, we verify that there is no correlation of the residuals with the parameters $m_{\star}$ and $x_{\mathrm{sat}}^{\mathrm{2D}}$, for neither TNG300 nor TNG100 galaxies.

    \subsection{Redshift evolution}
    \label{sec:zevol_obs}

Until now, we focused on galaxy clusters at redshift zero, while in most surveys, observed clusters are at higher redshifts. In this section, we check whether there is an evolution of the SsHMR with cosmic time, and give a parameterization of the SsHMR for galaxies in higher redshift clusters. For this, we perform the same measurement as presented in the previous sections, that is the full SsHMR for passive/active/all galaxies, and the radially dependant SsHMR for passive galaxies at redshift $z = 0.24$ and $z = 0.5$.

As cluster galaxies are getting quenched over time, we expect to have a larger fraction of active galaxies at higher redshift. Indeed, we give in Table~\ref{tab:shmr} the fractions of satellites that are quenched. It varies from 83\% at redshift $z=0$ to 76\% at $z=0.24$, and 71\% at $z=0.5$.  There is still a majority of quenched galaxies in clusters at that redshift, in agreement with observational measurements \citep{hennig2017}.

We fit Eq. \ref{equ:SsHMR} to the full, active and passive galaxy samples at redshift $z=0.24$ and $z=0.5$, and Equation \ref{equ:SsHMR_rsat} to the passive galaxy sample. We give the best fit parameters, and the 68\% confidence intervals for the different samples in Table~\ref{tab:shmr}.
When looking at each sample of galaxies individually (active or passive), there appears to be no significant time evolution of the best parameters. The small changes over time in the best fit parameters for the full galaxy population may therefore be attributed to the fluctuation of the passive to active galaxy fraction between the different redshift considered.

\section{Obtaining $m_{\star}$ from $m_{\mathrm{sub}}$}
\label{sec:shmr_simu}

    \subsection{Full subhalo sample}
    
Conversely, it can be useful to determine the distribution in stellar mass of cluster galaxies, given the distribution of their subhalo masses, for instance when populating a dark matter only simulation with galaxies. In this section, we give fitting functions for the median stellar mass of satellite galaxies as a function of subhalo mass, and for the scatter around this median relation. In this case, we select the galaxy sample to be complete in subhalo mass, and keep galaxies with $m_{\mathrm{sub}} > 10^{10} h^{-1}M_{\odot}$ et $0 < R_{\mathrm{sat}}^{\mathrm{3D}}/R_{200} < 2$.  
    
We first measure the SsHMR for all the selected subhaloes, similarly to what is presented in \citetalias{niemiec2019}, but updated for TNG300. We fit the SHMR function presented in \citet{moster2013}, which is the "inverse" function of the one discussed in Sect.~\ref{sec:full_shmr_obs}:

\begin{equation}
    m_{\star} (m_{\mathrm{sub}}) = 2N'\left[ \left(\frac{m_{\mathrm{sub}}}{M_1'}\right)^{-\beta'} + \left(\frac{m_{\mathrm{sub}}}{M_1'}\right)^{\gamma'}\right]^{-1}m_{\mathrm{sub}}.
    \label{equ:shmr_simu}
\end{equation}

\begin{figure} 
  \begin{center}
    \includegraphics{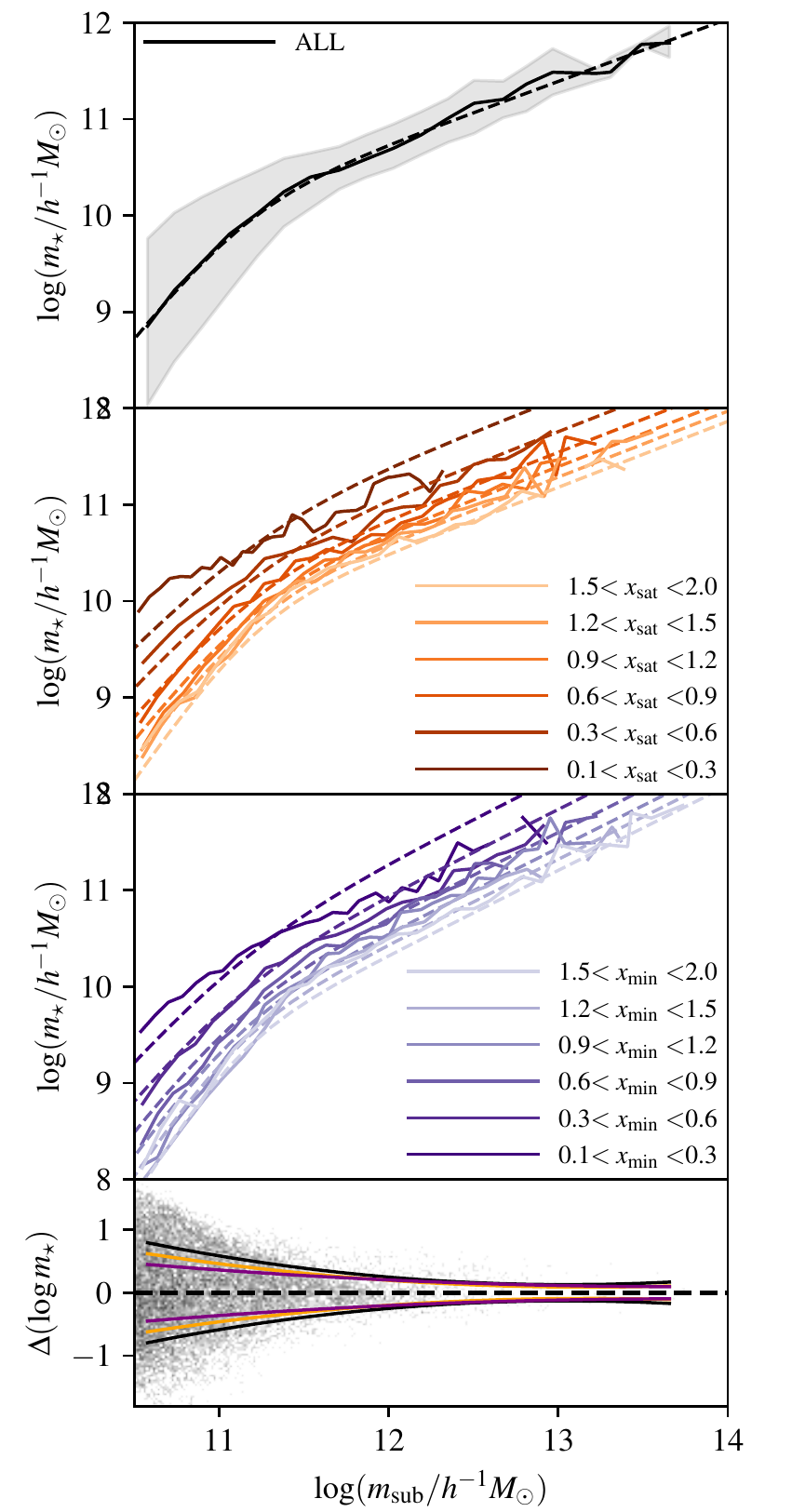}
    \caption{\textit{Top panel}: ShHMR measured for the satellite galaxies with $m_{\mathrm{sub}} > 10^{10}h^{-1}M_{\odot}$. The solid black line represents the median measured relation, and the grey shaded region the $16^{th}-84^{th}$ percentiles. The best fit relation corresponding to equation~\ref{equ:shmr_simu} is shown as a dashed line. \textit{Middle top panel}: SsHMR in bins in $x_{\mathrm{sat}}^{\mathrm{3D}}$, with solid lines showing the median relations, and dashed lines the best fit relations according to Eq. \ref{equ:shmr_simu_Rsat}. \textit{Bottom middle panel}: same but using $x_{\mathrm{min}}$ instead of $x_{\mathrm{sat}}^{\mathrm{3D}}$. \textit{Bottom panel:} residual distribution $\Delta(\log m_{\star}) = \log m_{\star} - \log m_{\star}(m_{\mathrm{sub}})$ around the best fit relation, with the parameterization from Equation~\ref{equ:shmr_simu_res} shown a red solid line. }
		\label{fig:shmr_simu}
  \end{center}
\end{figure}

Top panel of Fig.~\ref{fig:shmr_simu} presents the measured SsHMR, with the median relation shown as a solid black line, and the 16th-84th percentile region as a grey shaded area. The best fit parameters are given in Table~\ref{tab:shmr}, and the corresponding relation is shown as a dashed line. The full residual distribution is well fitted by a Gaussian function, with a RMS width $\bar{\sigma} = 0.59$. However, the width of the residual distribution has a strong dependence on the subhalo mass $m_{\mathrm{sub}}$, unlike Sect.~\ref{sec:full_shmr_obs} when the scatter width was fairly independent of $m_{\star}$. Bottom panel of Fig.~\ref{fig:shmr_simu} shows the residual distribution as a function of $m_{\mathrm{sub}}$, and the parametrization of the scatter is given as a function of $m_{\mathrm{sub}}$. To obtain that, we measure the half-width of the 16$^{th}$-84$^{th}$ percentile region of the residual distribution as a function of $m_{\mathrm{sub}}$, and fit this with a second order polynomial:
\begin{equation}
    \sigma (m_{\mathrm{sub}}) = 0.11\times m_{\mathrm{sub}}^2 -2.84 \times m_{\mathrm{sub}} + 18.67
    \label{equ:shmr_simu_res}
\end{equation}
This measurement of the scatter width is shown as a solid black line in the bottom panel of Fig.~\ref{fig:shmr_simu}.

    \subsection{Dependence on a secondary parameter}

As the scatter in the SsHMR is quite high, especially on the low mass end, we add as in Sect.~\ref{sec:shmr_obs} an extra parameter in Eq.~\ref{equ:shmr_simu}. The goal is to predict satellite galaxy stellar masses, given some parameters that could be for instance extracted from a N-body simulation. The easiest parameter to measure is the 3D distance between subhaloes and the centre of their host, normalized by the virial radius of the host, $x_{\mathrm{sat}}^{\mathrm{3D}}\equiv R_{\mathrm{sat}}^{\mathrm{3D}}/R_{200}$. We therefore first add a dependence on $x_{\mathrm{sat}}^{\mathrm{3D}}$ to the SsHMR in the following way: (1) the evolution of the SsHMR normalization with $x_{\mathrm{sat}}^{\mathrm{3D}}$ does not appear linear as in the case of Equation \ref{equ:SsHMR_rsat}, we include a power law dependence on $x_{\mathrm{sat}}^{\mathrm{3D}}$; (2) the slope of the low mass end power low, $\beta'$, also depends linearly on $x_{\mathrm{sat}}^{\mathrm{3D}}$. This gives:
\begin{equation}
    m_{\star} (m_{\mathrm{sub}}) = 2N'\left[ \left(\frac{m_{\mathrm{sub}}}{M_1'}\right)^{-\beta'} + \left(\frac{m_{\mathrm{sub}}}{M_1'}\right)^{\gamma'}\right]^{-1}m_{\mathrm{sub}}\times x_{\mathrm{sat}}^A,
    \label{equ:shmr_simu_Rsat}   
\end{equation}
with $\beta' = c*x_{\mathrm{sat}}^{\mathrm{3D}} + d$. 

We fit this function to the data and give the best fit parameters in Table~\ref{tab:shmr}. We show the SsHMR in bins of $x_{\mathrm{sat}}^{\mathrm{3D}}$ as solid lines in the top middle panel of Fig.~\ref{fig:shmr_simu}, as well as the best fit functions in the same bins as dashed lines. We measure the scatter around the best fit, and parametrize it again as a function of $m_{\mathrm{sub}}$ with a second order polynomial, which gives: 
\begin{equation}
    \sigma_{\mathrm{x_{sat}}} = 0.08\times m_{\mathrm{sub}}^2 -2.08 \times m_{\mathrm{sub}} + 13.79.
\end{equation}
This parametrization of the scatter is shown with orange solid lines in the bottom panel of Fig.~\ref{fig:shmr_simu}.
Including the cluster-centric distance improves the prediction of the stellar mass for a  given subhalo mass: the overall scatter is reduced to $\bar{\sigma}_{\mathrm{x_{sat}}} = 0.45$,   as compared to the full SsHMR scatter $\bar{\sigma} = 0.59$.

To further reduce this scatter, it is necessary to examine parameters that can better trace the orbital history of subhaloes within their host. Indeed, although the cluster-centric distance at redshift zero is correlated with the time of accretion, the degree to which a subhalo is stripped (and possibly the degree to which the corresponding galaxy is quenched) also depends on the shape of the orbit that the subhalo had followed during infall. To take this effect into account, we chose as a secondary alternative parameter, 
the distance of minimal approach to the host centre normalized by the host virial radius: $x_{\mathrm{min}} = \min(x_{\mathrm{sat}}^{\mathrm{3D}})_{tacc \rightarrow t}/R_{200}$. We include this parameter in the SsHMR also following Eq. \ref{equ:shmr_simu_Rsat} and fit again the parameters $N',\, M_1',\, \gamma',\, A,\, c,\, d$. The best fit parameters are given in Table~\ref{tab:shmr}, and the SsHMR measured in $x_{\mathrm{min}}$ bins is shown as solid lines in the bottom middle panel of Fig.~\ref{fig:shmr_simu} along with the best fit function  in the same bins, shown with dashed lines.

We fit the residuals along the best fit relation with a second order polynomial, giving an expression for the scatter:
\begin{equation}
    \sigma_{\mathrm{x_{min}}} = 0.04\times m_{\mathrm{sub}}^2 -0.97 \times m_{\mathrm{sub}} + 6.76,
\end{equation}
which is shown with a solid purple line on the bottom panel of Fig.~\ref{fig:shmr_simu}. Using $x_{\mathrm{min}}$ instead of $x_{\mathrm{sat}}^{\mathrm{3D}}$ allows to further reduce the scatter, with a total value of $\bar{\sigma_{\mathrm{x_{min}}}} = 0.34$. We chose to present both fits with $x_{\mathrm{sat}}^{\mathrm{3D}}$ and $x_{\mathrm{min}}$, as $x_{\mathrm{sat}}^{\mathrm{3D}}$ is quite straightforward to obtain from a simulation snapshot, while $x_{\mathrm{min}}$ requires to have access to not only one snapshot, but to the whole merger trees and orbital info of subhaloes. Therefore, depending on the application, it can be more useful to parametrize the SsHMR as a function of either the cluster-centric distance at redshift zero, or the distance of minimum approach. Still, as seen in the decrease of scatter when considering $x_{\mathrm{min}}$ instead of $x_{\mathrm{sat}}^{\mathrm{3D}}$, the instantaneous position in the cluster has less impact on the SsHMR than the orbital history.

We note that there appears to be a discrepancy in Fig.~\ref{fig:shmr_simu} between the best fit function and the measurement at the high mass end for galaxies within the inner most bin (both in terms of $x_{\mathrm{sat}}^{\mathrm{3D}}$ and $x_{\mathrm{min}}$). However, this concerns only a very small number of galaxies: there are only 12 (73) galaxies over a total of $\sim 19 000$ that have $m_{\mathrm{sub}} > 10^{12}h^{-1}M_{\odot}$ and $x_{\mathrm{sat}}^{\mathrm{3D}} < 0.3$ ($x_{\mathrm{min}} < 0.3$).

    \subsection{Redshift evolution}

As in Sect.~\ref{sec:zevol_obs}, we now give fitting formulae for the SsHMR measured at higher redshift, as it can be useful to populate subhaloes with galaxies for clusters at earlier times. We follow the same procedure as at redshift zero, and fit $m_{\star} = f(m_{\mathrm{sub}})$ as given in Equation \ref{equ:shmr_simu} for subhaloes/satellite galaxies at redshift $z = 0.24$ and $z=0.5$. The best fit values, along with the overall scatter, are given in Table~\ref{tab:shmr}. We note that there is no significant evolution of the best fit parameters within the studied redshift range.

We also include the dependence on the cluster-centric distance $x_{\mathrm{sat}}^{\mathrm{3D}}$ as defined in Equation \ref{equ:shmr_simu_Rsat}. We fit this for galaxies at redshift $z=0.24$ and $z=0.5$, and give the best fit parameters in Table~\ref{tab:shmr}. Again, there is no significant evolution within the considered redshift range.

\section{Time evolution of satellites properties}
\label{sec:time_evol}

    \subsection{All satellites}
    \label{sec:time_evol_all}
 
In this section, we examine the processes that lead to the shift in SsHMR for satellite galaxies compared to centrals, and are responsible for the scatter in the SsHMR. We extract the evolution of the different subhalo/satellite galaxy properties from simulation merger trees, starting at the time of their first crossing of the accretion radius defined here as $R_{
\mathrm{acc}} = 2 \times R_{200}$. We select satellite galaxies as in Sect.~\ref{sec:shmr_obs}, i.e. with $m_{\star} > 10^{9}h^{-1}M_{\odot}$ and size $> 2h^{-1}\rm{kpc}$.

\begin{figure} 
  \begin{center}
    \includegraphics{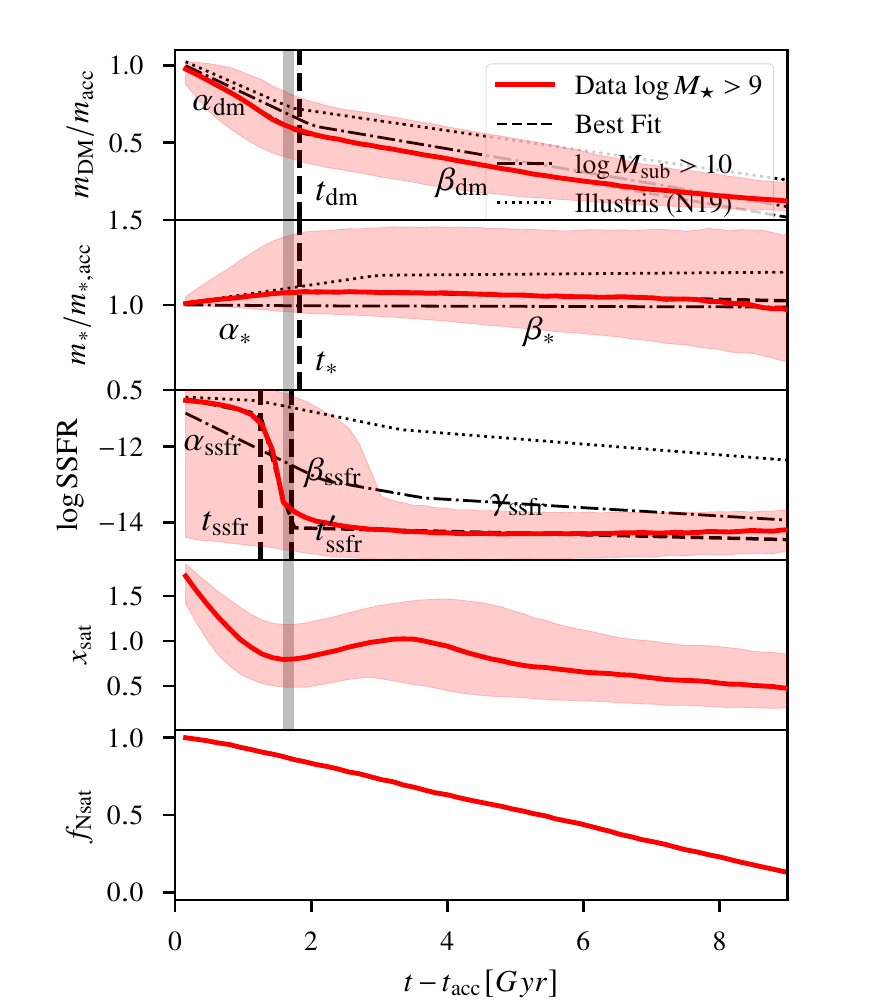}
    \caption{Median evolution of $m_{\mathrm{DM}}/m_{\mathrm{acc}}$, $m_{\mathrm{*}}/m_{\mathrm{*, acc}}$, SSFR, distance to the cluster centre normalized by $R_{200}$ at accretion, and number of galaxies at each time step (from top to bottom) as a function of time (red line, with
      the 16th-84th percentiles shown as red region), and the best fit
      evolution (black dashed lines), for satellite galaxies with $m_{\star} > 10^{9}h^{-1}M_{\odot}$ at redshift $z=0$. The grey vertical line
      represents the time at which the mean evolution of the
      cluster-centric distance reaches its first minimum. The dash-dotted lines show the evolution when selecting subhaloes with $m_{\mathrm{sub}} > 10^{10}h^{-1}M_{\odot}$ at redshift $z=0$, and the dotted lines show the same evolution as measured in \citetalias{niemiec2019} for the Illustris simulation.}
		\label{fig:tevol_fit}
  \end{center}
\end{figure}

Figure~\ref{fig:tevol_fit} shows the time evolution of the different subhalo/galaxy properties since the time of infall, from top to bottom, respectively: the dark matter mass of the subhaloes normalized by mass at the time of accretion, the stellar mass normalized by mass at accretion, the specific star formation rate, and the distance to the centre of the host normalized by $R_{200}$ at time of accretion. The red line shows the median evolution for all subhaloes, and the shaded area the 16th-84th percentiles. We fit the different evolutions, represented as dashed lines.

The time evolution of satellite properties was the main focus of the study presented in \citetalias{niemiec2019}, but it was done with the previous version of the Illustris simulation. As a comparison, we also show the evolution obtained when selecting subhaloes as in \citetalias{niemiec2019} with $m_{\mathrm{sub}} > 10^{10}h^{-1}M_{\odot}$ in dot-dashed lines,  and the  evolution as measured for the Illustris simulation in \citetalias{niemiec2019} with dotted lines. We discuss the difference between the two simulations in Sect.~\ref{sec:diffs_ill}.

We focus here on galaxies selected in stellar mass (red solid line + dashed black):
in a similar way as in \citetalias{niemiec2019}, the main evolution is driven by a decrease in the dark matter subhalo mass due to tidal stripping. The dark matter loss rate is stronger in the phase of first infall. This phase is defined as the time between the first crossing of the accretion radius, and the time that the median evolution of the cluster-centric distance reaches its first minima (i.e first pericentre in the satellite galaxies orbit within the cluster). This first closest approach is shown in Fig.~\ref{fig:tevol_fit} as a grey bar, and happens at $t_{\mathrm{peri}} = t_{\mathrm{acc}} + 1.67 Gyr$.

We fit the following function to the dark matter mass evolution:
\begin{equation}
	\frac{m_{\rm{DM}}}{m_{\rm{acc}}}(t) = \begin{cases} 
    	\alpha_{\rm{dm}}t + c_{\rm{dm}}, & \mbox{if } t < t_{\rm{dm}} \\ 
    	\beta_{\rm{dm}}t + c'_{\rm{dm}}, & \mbox{if } t > t_{\rm{dm}}, \end{cases}
        \label{equ:broken_line}
\end{equation}
where $\alpha_{\mathrm{dm}}$ and $\beta_{\mathrm{dm}}$ are the slopes of the evolution, and $t_{\mathrm{dm}}$ is the time of slope change. We give the best fit values for the parameters in Table~\ref{tab:fit_evol}. During the phase of first infall, subhaloes lose on average $\sim 25\%$ of their mass per Gyr, and the mass loss rate goes down to $\sim 7\%$ per Gyr after the first pericentre crossing. We note that some galaxies can be stripped of up to 90\% of their dark matter mass, but  not all galaxies reach this stage. To quantify that, we compute the surviving DM mass fraction for subhaloes as $f_{\mathrm{surv}}^{\mathrm{DM}} = m_{\mathrm{DM}}(z = 0)/m_{\mathrm{DM}}(z_{\mathrm{acc}})$. Only 10\% of subhaloes end up with less that 10\% of their mass at accretion at $z = 0$, while 51\% retain between 10 and 50\% of their mass at accretion, and 37\% between 50 and 100\%. 2\% of subhaloes even continue to gain mass after accretion and end up with a higher mass than at accretion, possibly due to mergers (with other subhaloes) in the dense cluster environment.

As for the baryonic component of the galaxies, the median stellar mass only increases by $\sim7\%$ of the mass at accretion during the phase of first infall, before star formation is quenched on average. After this first phase, the stellar mass even start to decrease. The apparent loss in stellar mass is due partly to the low number of galaxies that spend the whole time range in the cluster (from $t_{\mathrm{acc}}$ to $t_{\mathrm{acc}} + 9Gyr$). To quantify how much stellar mass is actually lost, we compute again the surviving mass fraction, but for the stellar component of the galaxies: $f_{\mathrm{surv}}^{\star} = m_{\star}(z = 0)/m_{\star}(z_{\mathrm{acc}})$. We find that 60\% of galaxies have a lower stellar mass at redshift $z=0$ than at accretion, but this mass loss remains lower than for the dark matter component, as 70\% of these galaxies lose less than 20\% of their mass at accretion. We note that the amount of stellar mass loss is dependent on the simulation resolution (see Sect.~\ref{sec:discussion}), and galaxies in TNG100 end up with lower stellar surviving fractions than in TN300, but still much higher than the dark matter surviving fraction.  This is consistent with tidal stripping of part of the stars, which happens only if a significant amount of dark matter is initially stripped \citep{smith2016}: galaxies that have lost stellar mass during infall are the ones that lost a higher amount of dark matter mass ($< f_{\mathrm{surv}}^{\mathrm{DM}} > = 40$), compared to galaxies that have gained (or conserved) stellar mass ($< f_{\mathrm{surv}}^{\mathrm{DM}} > = 70$).
It is important to note that a large fraction of galaxies does not experience stripping of their stellar component: if the median stellar mass decreases, it is partly due to the fact that we do not follow all satellite galaxies during the whole time range, as some spend much less than 9 Gyr in the cluster. Bottom panel of Fig.~\ref{fig:tevol_fit} shows the number of galaxies at each time step (relative to the total number), and for instance only half of the satellites at redshift $z=0$ spend at least $\sim 5 Gyr$ in the cluster. 
We also fit Eq.~\ref{equ:broken_line} to the stellar mass evolution, and give the best fit parameters in Table~\ref{tab:fit_evol}. We note that the scatter is high for the evolution of the stellar component, some galaxies may form star at a high rate during their first infall into the cluster (for instance with ram-pressure induced starburst).

The evolution of the specific star formation rate confirms this scenario: at accretion, galaxies are on average still forming stars, and after $\sim1.2\,$Gyr there is a rapid transition, lasting $\Delta t_{\mathrm{ssfr}}\sim 0.5\,$Gyr, into a population of on average quenched galaxies. Compared to what we measured in \citetalias{niemiec2019} for the Illustris model, the transition is shorter and starts earlier, producing a population of mainly quenched galaxies at the time of first pericentre crossing. In \citet{wetzel2013}, they combined a group and cluster catalogue from SDSS DR7 with N-body simulation, and find a similar quenching time-scale, $\Delta t_{\mathrm{ssfr}} < 0.8\,Gyr$, but a much longer delay before the transition onset, lasting $2-4\,Gyr$. However, they consider as time of infall the \emph{first infall} into any groups or clusters, therefore taking into account preprocessing into the satellite SFR evolution (see Sect.~\ref{sec:preprocessing} for a discussion of the impact of preprocessing in the evolution measured here).  As many galaxies have unresolved SFRs and have been attributed random values (see Sect.~\ref{sec:sample}), the median value of the quenched population (after transition) may not be completely significant. Instead of giving the full parameterization of this evolution, we simply give the time of the start of the transition, and the time of the end, $t_{\mathrm{ssfr}} = 1.25 \pm 0.03$ Gyr and $t'_{\mathrm{ssfr}} = 1.71 \pm 0.11$ Gyr.

As a first step, it is interesting to look at the median evolution of these properties, but this does not give much handle on the scatter on the observed SsHMR at redshift $z=0$, except by the fact that not all galaxies experience the full evolution for 9\,Gyr (as shown in the bottom panel of Fig.~\ref{fig:tevol_fit}). To better understand how different galaxies are affected in different proportions by main mechanisms in clusters (tidal stripping of dark matter, and quenching vs star forming), we need to look in more details at the time evolution for different galaxy sub-samples.

\begin{table}
\centering
\begin{tabular}{|c  | c c | }
\multicolumn{3}{c}{All satellites} \\
\hline
		& $m_{\rm{DM}}$			& $m_*$				\\
\hline
$\alpha$& $-0.253 \pm 0.003$	& $0.042 \pm 0.001$\\
$\beta$	& $-0.076 \pm 0.001$	& $-0.008 \pm 0.001$\\
$c$		& $1.025 \pm 0.003$		& $1.005 \pm 0.001$	\\
$t$	    & $1.83 \pm 0.02$		& $1.83 \pm 0.03$	\\
\hline
\multicolumn{3}{c}{}\\
\multicolumn{3}{c}{Passive satellites} \\
\hline
		& $m_{\rm{DM}}$			& $m_*$				\\
\hline
$\alpha$& $-0.271 \pm 0.004$	& $0.034 \pm 0.001$	\\
$\beta$	& $-0.077 \pm 0.001$	& $-0.010 \pm 0.001$\\
$c$		& $1.022 \pm 0.004$		& $1.008 \pm 0.001$	\\
$t$	    & $1.72 \pm 0.35$		& $1.86 \pm 0.04$	\\
\hline
\end{tabular}
\caption{Best fit parameters of the evolution of the dark matter and
  stellar masses, as a function of
  time, as shown in Fig.~\ref{fig:tevol_fit}. The fits are performed on the median evolution over all
  satellite galaxies (top table), and only satellites quenched at $z = 0$. }
\label{tab:fit_evol}
\end{table}

    \subsection{Influence of galaxy properties}
    \label{sec:tevol_galprop}

\begin{figure*} 
  \begin{center}
    \includegraphics{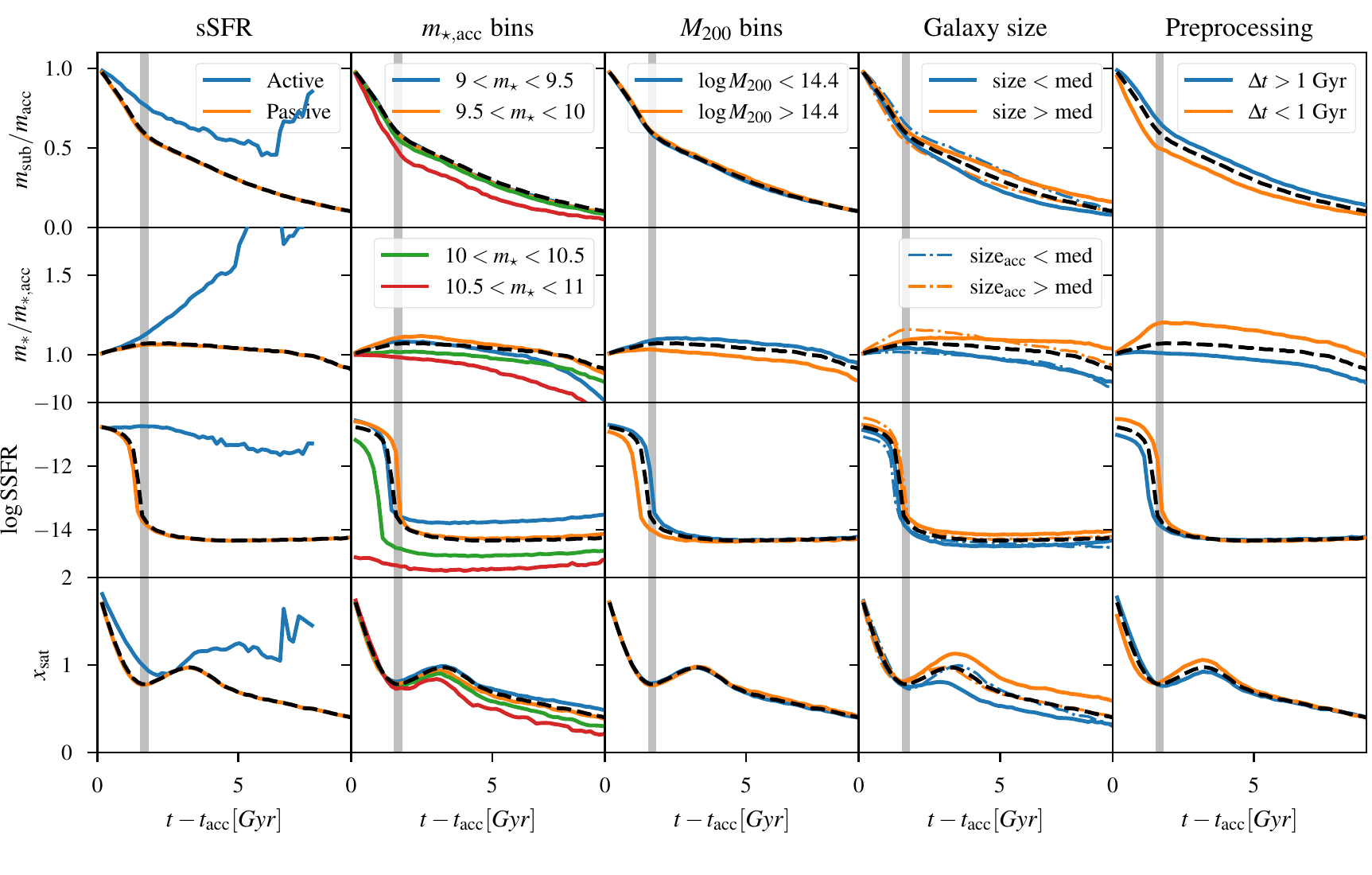}
    \caption{Median evolution of $m_{\mathrm{sub}}/m_{\mathrm{acc}}$
      (\textit{top panel}), $m_{\mathrm{*}}/m_{\mathrm{*, acc}}$ (\textit{top middle panel}),  SSFR (\textit{bottom middle panel}) and $x_{\mathrm{sat}}^{\mathrm{3D}}$ (\textit{bottom panel}), as a function of time. The different sub-population of galaxies that are considered are passive vs active at $z=0$ (\textit{left panel}), bins in $m_{\star\mathrm{,acc}}$ (\textit{middle left panel}), high vs low host cluster mass (\textit{middle panel}), galaxy size at given stellar mass (\textit{middle right panel}) and preprocessed or not galaxies (\textit{right panel}).}
		\label{fig:tevol_alldifs}
  \end{center}
\end{figure*}

We now examine the evolution of  different sub-populations of satellite galaxies (unlike \citetalias{niemiec2019}, where we only looked at the evolution of all galaxies together). The time evolution of galaxy properties  in different sub-samples is shown in Fig.~\ref{fig:tevol_alldifs}.

We first consider the time evolution of galaxies that are still star-forming at redshift $z=0$ as compared to galaxies that are quenched (blue and orange respectively in the left panel of Fig.~\ref{fig:tevol_alldifs}). Active galaxies represent a small sub-population at redshift $z=0$ (18\%), so the evolution of passive galaxies drive the global evolution. Galaxies that are still active at redshift $z=0$ were continuously forming star during their time spent in the cluster, and present a drastically different $m_{\star}$ evolution than the passive population. In addition, these galaxies have been less affected by the cluster dense environment, and therefore less subject to tidal stripping of the dark matter component. As for the reasons these galaxies were less affected by their host cluster, there are two explanations that we can deduce from our measurements: (1) these galaxies have spent less time in the cluster environment (see Sect.~\ref{sec:full_shmr_obs}), and (2) they appear to be on average on a different type of orbits than the passive population, with a less rapid infall, and a pericentre further from the cluster centre (i.e. larger $x_{\mathrm{min}}$).

We then examine the influence of the stellar mass at the time of accretion on galaxy evolution (middle left panel of Fig.~\ref{fig:tevol_alldifs}). As could be expected, higher mass galaxies infall deeper into the centre of their host due to dynamical friction, and therefore lose a higher fraction of their mass. More massive galaxies are quenched faster, and the most massive ones are already quenched when they start their infall. Indeed, most massive galaxies are prone to quenching by AGN feedback (i.e intrinsic quenching) rather than by interactions with the environment \citep[see for instance][]{donnari2021a}. 

We also look at the impact of the host halo mass on galaxy evolution as shown in the middle panel of Fig.~\ref{fig:tevol_alldifs}. As cluster size is absorbed into the cluster-centric distance definition ($x_{\mathrm{sat}}^{\mathrm{3D}} \equiv R_{\mathrm{sat}}^{\mathrm{3D}}/R_{200}$), there is no impact on the orbital evolution. More interestingly, the amount of dark matter mass loss does not appear to be affected by the mass of the host halo. This could seem counter-intuitive at a first glance, as for a subhalo at a given distance $x_{\mathrm{sat}}^{\mathrm{3D}}$ from its host cluster,  tidal forces are proportional to the cluster mass enclosed within the  distance $x_{\mathrm{sat}}^{\mathrm{3D}}$: more massive halos are denser in the core, and should therefore exert stronger tidal forces, and lead to higher amount of stripping. This picture however omits the fact that subhalo orbits also scale with the host cluster masses, and when looking at \emph{absolute distances}, subhaloes infall closer to the centre of less massive hosts. The self-similarity in subhalo orbits when scaled by $R_{200}$, and in the resulting dark matter loss, is still an important feature to note. \citet{engler2021} showed that in the TNG simulation, this can be extrapolated to subhaloes residing in lower mass groups (with $M_{200} > 10^{12}M_{\odot}$), that exhibit the same amount of dark matter mass loss as subhaloes residing in clusters.
However, for the baryonic component, galaxies located in higher mass clusters are quenched faster. Using a simple analytical model, \citet{hester2006} showed that the amount of ram-pressure stripping of the gas contained in a satellite galaxy's outer H1 disk and hot galactic halo depends on the ratio between the satellite and the host total mass; as the stellar mass distribution of galaxies residing in the low and high mass cluster bins does not present significant differences, this model would predict a different amount of stripping in both samples, potentially resulting in different star-formation evolutions. In addition, part of the difference could be explained by the larger amount of preprocessed galaxies in more massive clusters in the TNG simulation, as shown in \citet{donnari2021a}. This is consistent with a picture where the subhalo mass function does not depend on the host halo mass, while more massive clusters contain a higher fraction of quenched galaxies.

Finally, we examine the impact of the galaxy "compactness" on its evolution, as this parameter has shown to have an influence on the SsHMR (see Sect.~\ref{sec:shmr_obs_params}).  For this, we measure the median galaxy size in galaxy stellar mass bins, and define the large (small) galaxy sample as having their size larger (smaller) than the median size in their stellar mass bin. The two galaxy samples appear to have very different evolutions (middle right panel of Fig.~\ref{fig:tevol_alldifs}): more compact galaxies infall deeper into their host, and therefore lose a higher fraction of their dark matter mass and get quenched faster. To verify if this difference in size is simply a result of two galaxy populations that happen to be on different orbits and therefore are differently affected by the cluster, which in turns results in different sizes at redshift $z=0$, or if the compactness of galaxies at accretion sets them on different evolutionary paths, we measure again the time evolution but splitting galaxies with respect to their size \emph{at accretion}. These evolutions are shown in the same panel of Fig.~\ref{fig:tevol_alldifs} with dash-dotted lines. When selecting galaxies on their size at accretion (instead of size at redshift 0), trends in evolution of stellar mass (or sSFR) are the same, but amplified: more compact galaxies appear to be quenched faster, and, on average, do not form stars during their infall, while extended galaxies are less easily quenched. 
What is even more interesting, is that the evolution of the dark matter component shows an opposite trend: galaxies that are more compact at accretion are less prone to tidal stripping than extended ones, while at the same time having very similar median orbital histories. This could be explained if the concentration of galaxies is correlated with the concentration of their host subhaloes at accretion. Then, at a given orbit and a given mass, extended subhaloes would be more easily stripped than more concentrated ones.

These two evolutionary paths (with respect to galaxy size at accretion or at redshift $z=0$), give a more complete picture of the impact of galaxy compactness. Galaxy concentration at accretion, which is not an observable in real data-sets, partly drives the co-evolution of galaxies and their subhaloes. On the other hand, galaxy compactness at redshift $z=0$, which can potentially be measured in observational data, results from both the compactness at accretion but also from the impact of stellar stripping during infall. It is therefore a parameter to consider, and study in more details when trying to measure a tidal stripping signal in observational analyses, for instance using gravitational lensing.

    \subsection{Influence of preprocessing}
    \label{sec:preprocessing}

A significant fraction of galaxies that fall into a cluster were already satellites in smaller groups before, and were therefore subjected to previous environmental interactions that modified their properties, which is known as preprocessing \citep{mcgee2009, bahe2013, hou2014}. In the TNG simulation, $\sim 30\%$ of cluster satellites were  already quenched before their accretion into their redshift 0 host, and for low mass galaxies this can be mainly attributed to preprocessing in groups with mass higher than $10^{12}M_{\odot}$, as shown in  \citet{donnari2021a}.
Here we consider the impact of preprocessing on the co-evolution of satellite dark and baryonic matter, and chose a definition of preprocessed galaxies that does not rely on their merger history: we consider a galaxy as being preprocessed if its mass has been gravitationally affected prior to its infall into its redshift 0 host. Indeed, central galaxies are expected to increase their total mass over time, by continuous matter accretion and by mergers. On the other hand, halo mass loss is mostly due to gravitational interaction with other (sub)haloes, such as stripping by tidal forces. A subhalo therefore reaches its maximum mass, $m_{\mathrm{sub}}^{\mathrm{peak}}$, at the transition between the accreting and mass-loss phases of this evolution, and this maximum mass been shown to correlate well with satellite galaxy properties \citep[e.g.][]{reddick2013}. 
Following this logic, we define preprocessed galaxies as the ones having reached their maximal subhalo mass \emph{before their accretion into their host cluster}, and that have therefore started losing mass before being satellites of their current host, for instance by tidal stripping in groups \citep{joshi2019}. We chose as threshold $\Delta t = m_{\mathrm{sub}}^{\mathrm{peak}} - m_{\mathrm{sub}}^{\mathrm{acc}} = 1\mathrm{Gyr}$.

Right panel of Fig.~\ref{fig:tevol_alldifs} shows the time evolution of preprocessed (\emph{blue}) and not preprocessed (\emph{orange}) galaxies. Preprocessed galaxies lose a smaller fraction of their subhalo mass during their accretion into the cluster. A possible explanation for that is that these galaxies have already been stripped of the outskirts mass  during their interaction within their previous host, and therefore the remaining mass is more gravitationally bound.

Also, the evolution of the stellar mass for these two population is sensibly different: as could be expected, preprocessed galaxies have been quenched during their former interaction, and thus do not form any new stars during their infall into the cluster on average, while not preprocessed galaxies continue to form star during their first infall, leading to a median increase of 20\% of their stellar mass. This shows that the time of maximum mass is a good proxy for the time of quenching. We note that although galaxies can lose part of their dark matter and have their star formation stopped due to preprocessing, there is (almost) no stripping of the stellar component during this phase: 85\% of preprocessed galaxies have a larger stellar mass at accretion than at the time of maximum subhalo mass. The different evolution in stellar and dark matter masses for these two samples of galaxies shows that preprocessing plays a big part in the SsHMR scatter. \citet{rhee2017} reached similar conclusions, however this is not an easy parameter to derive from observations.

\section{Discussion}
\label{sec:discussion}

    \subsection{Differences between Illustris versions}
    \label{sec:diffs_ill}

The TNG simulation has brought improvement in the modeling of some key physical processes compared to the initial Illustris simulation, in order to better reproduce a large variety of observables in the Universe. Therefore, checking how the Illustris model for the satellite/subhalo evolution compares to TNG can lead to some understandings on how and which baryonic processes drive the property distribution of cluster galaxies. 
To make that comparison, we measure again the time evolution of satellite galaxy properties, but this time by applying the same selection criteria as in \citetalias{niemiec2019}, i.e. keeping galaxies with $m_{\mathrm{sub}} > 10^{10}h^{-1}M_{\odot}$. This selection is the same as in Sect.~\ref{sec:shmr_simu}. We show in Fig.~\ref{fig:tevol_fit} this evolution measured in TNG with dot-dashed lines, and the one measured in Illustris by \citetalias{niemiec2019} with dotted lines.

The most dramatic difference between satellite galaxies evolution in TNG vs Illustris is that of the stellar mass: in Illustris, galaxies continue on average to form stars during more than 2Gyrs after accretion, leading to an average increase of 20\% of their mass at accretion. In TNG, on the contrary, when applying the same selection as by \citetalias{niemiec2019}, the stellar mass evolution is flat. Even if a fraction of galaxies continue to form stars after accretion or even experience a starburst episode induces by ram-pressure or tidal interactions \citep{lotz2018}, the median evolution of the stellar mass over the whole galaxy sample does not show any increase. There are two possible explanations to this difference: either the different implementations of galaxy evolution models lead to a different strength of quenching mechanisms \emph{in the clusters}, or different processes affect the star-formation of galaxies \emph{before} they are accreted (some mixture of both is also possible). 

To quantify the contribution of in-cluster quenching vs pre-cluster quenching in both versions of the simulation, we compute the fraction of galaxies that are quenched at their time of infall and at redshift $z=0$, for both the Illustris-1 and TNG300 simulations. At the time of accretion at $2\times R_{200}$, 52\% of galaxies with $m_{\mathrm{sub}} > 10^{10} h^{-1}M_{\odot}$ are still forming stars in the Illustris-1 simulation, while only 40\% in the TNG simulation. This can be due in part to the new implementation of AGN feedback, that has been shown to quench more galaxies in haloes with masses in the range $10^{12}-10^{14}M_{\odot}$ than  in Illustris \citep{weinberger2018}. Overall, the new implementation of galactic winds has shown to reduce the star-formation rates and thus stellar masses of galaxies at all mass scales \citep{pillepich2018a}. However, at redshift $z=0$, the difference in quenched ratios is even more dramatic: $\sim 30\%$ of galaxies included in the \citetalias{niemiec2019} sample are still forming stars, compared to only 6\% of satellite galaxies with $m_{\mathrm{sub}} > 10^{10} h^{-1}M_{\odot}$ in TNG300. Even if part of this difference is due to the overall shift in the SFR between galaxies in Illustris and in TNG, this could still suggest that the different implementations specifically impact some quenching processes that happen in clusters. \citet{donnari2021a} suggests that low mass satellite galaxies are less prone to ram-pressure stripping in Illustris than in TNG, as stellar feedback causes them to have a higher gas content \citep{pillepich2018a, diemer2019}, while groups are more deprived of their gas because of the strongly ejective AGN feedback implemented in Illustris \citep{pillepich2018a, terrazas2020}.

The second difference between the evolution in Illustris vs TNG concerns the amount of dark matter stripping, which appears more important in TNG than in Illustris. This can be considered surprising as tidal stripping is a purely gravitational interaction, and therefore should not depend on the implementation of subgrid baryonic processes. One possible explanation would be that the sample we consider here contains more massive clusters  in Illustris (in \citetalias{niemiec2019} only three clusters with masses $\log ( M_{200}/h^{-1}M_{\odot}) \sim 14.2$), but the evolution measured in $M_{200}$ bins in Sect.~\ref{sec:tevol_galprop} did not show any difference for stripping in more or less massive clusters.
We also verify that this is not due to resolution differences between TNG300-1 and Illustris-1, by measuring the time evolution of satellite properties in TNG100, which showed an amount of stripping consistent with what is found in TNG300. One possible explanation for this, is that galaxies have been shown to have higher stellar-to-halo mass fractions in Illustris compared to TNG. If there is more (stellar) mass in the centre of the subhalo, it can make the gravitational potential stronger or more concentrated, therefore making stripping harder. This is a good example of how baryonic processes can actually impact the distribution of dark matter \citep[see also][]{duffy2010, sorini2021}. Although there does not seem to be any strong differences between the galaxy samples in Illustris and TNG, it is also possible that the difference in stripping could be resulting from the small cluster sample considered in the Illustris analysis presented in \citetalias{niemiec2019} (only three haloes with $M_{200} > 10^{14}h^{-1}M_{\odot}$).

    \subsection{Numerical effects}

Numerical effects are a possible source of imprecision in results derived from numerical simulations. For instance, it has been shown \citep{pillepich2018a, weinberger2018} that in TNG, the stellar mass of galaxies is impacted by the resolution of the simulation: at a given halo mass, stellar mass is higher in more resolved runs of the simulation than in less resolved ones. This is what motivates the stellar mass correction that we applied as described in Sect.~\ref{sec:sample}, following prescriptions described for in instance in \citet{pillepich2018b} and \citet{engler2021}. We note that some studies such as \citet{engler2021} apply a further correction to obtain masses in agreement with the most resolved run TNG50. We do not apply this second correction as TNG50 does not contain clusters massive enough to extract satellite galaxies that correspond to the ones we study here. The stellar masses that we give can therefore be under-estimated compared to a more resolved simulation run.   

In addition to stellar masses, it is possible that resolution affects the physical processes that create the shift (and the scatter) in the SsHMR as compared to the one for central galaxies. To verify this, it is not enough to examine the difference in the SsHMR (or the stripping factors) between the TNG300 and the TNG100 runs, as all differences would be absorbed in the $m_{\star}|m_{\mathrm{sub}}$ correction. We will thus directly compare the time evolution of the different galaxy parameters, as extracted from  TNG300 vs TNG100 runs. To quantify this, we compute again the surviving mass fractions for subhaloes, defined as in Sect.~\ref{sec:time_evol_all}, $f_{\mathrm{surv}} = m_{\mathrm{DM}} (z = 0)/m_{\mathrm{DM}}(z_{\mathrm{acc}})$. At first glance, there does not seems to be a strong impact of the resolution on the amount of dark matter stripping, as the distributions of $f_{\mathrm{surv}}$ between the two simulations are very similar: 61\% and 64\% of the considered subhalo sample in the TNG300 and TNG100 runs respectively, end up  with less than half of their mass at accretion at redshift $z=0$. However, the differences between the two runs are mainly apparent for the most stripped galaxies in the samples: In TNG300, 10\% of the sample has a surviving mass fraction of less than 10\%, while it reaches 17\% of the sample in TNG100. This may be due to the artificial disruption of some heavily stripped subhaloes.

As for the baryonic component, the difference between the two runs is more striking, and not only quantitative but also qualitative. In TNG300, galaxies that lose matter during accretion represent 60\% of the sample, and are therefore only slightly dominant as compared to galaxies who have gained mass (forming stars for at least part of their accretion). On the contrary, in TNG100, a vast majority of galaxies (84\%) end up with lower stellar mas than at accretion. The $f_{\mathrm{surv}}^{\star}$ distributions are very different in the two runs of the simulation, with a peak at  $f_{\mathrm{surv}}^{\star} \sim 0.97$ and $\sim 0.75$ in TNG300 and TNG100 respectively. 

Finally, we check if the resolution affects the scatter in the SsHMR. We measure the residuals $ \log(m_{\mathrm{sub}}) - \log(m_{\mathrm{sub}}(m_{\star}))$, with respect to Eq.~\ref{equ:SsHMR}, but for the satellite galaxies taken from the more resolved TNG100 run. We found the mean of the residual distribution to be consistent with 0, but the scatter to be slightly lower than in TNG300 for the full and the passive galaxy samples ($\sigma_{\rm{all}}^{\rm{TNG100}} = 0.46$ and $\sigma_{\rm{passive}}^{\rm{TNG100}} = 0.42$). However, this does not give much indication on the impact of resolution on the scatter, as the smaller box size of the TNG100 run gives a limited coverage of the cluster parameter space compared to TNG300. We therefore measure again the residual distribution using satellite galaxies from the TNG100-2 run. As described in Sect.~\ref{sec:data}, this run has the same box size as TNG100, but the same resolution as TNG300. We found that the with of the residual distribution is sensibly the same in TNG100-2 ($\sigma_{\rm{all}}^{\rm{TNG100-2}} = 0.45$ and $\sigma_{\rm{passive}}^{\rm{TNG100-2}} = 0.43$) as in TNG100, indicating that the resolution does not have a significant impact on the SsHMR scatter, given the cuts we have applied on our galaxy selection. We further verify that adding a dependence on the cluster-centric radius $x_{\rm{sat}}$ as described in Sect.~\ref{sec:sshmr_rsat}, reduces the scatter in a similar way for satellites in the TNG100 and TNG100-2 runs.

\section{Conclusions}

In this paper, we scrutinize the Stellar-to-subHalo mass relation (SsHMR) for cluster galaxies in the TNG300 simulation, paying a special attention to the scatter in this relation, and the physical parameters than can be used to better understand and constrain this scatter. We analyze the SsHMR and its scatter in two complementary ways: on one hand the "observational" point of view, when we want to predict the subhalo mass of cluster galaxies given their stellar mass, and other observable parameters. In this, we find that star-forming and passive galaxies follow distinct SsHMRs, and the (large) scatter in the passive SsHMR is correlated with the projected 2D distance to the centre of the host cluster. We also find that the galaxy "compactness", defined as the galaxy size at a given stellar mass, also helps to predict subhalo masses, when also considering the stellar mass and cluster-centric distance. We give in Sect.~\ref{sec:shmr_obs} convenient fitting functions to predict subhalo masses (median + scatter) as a function of either the stellar mass alone, or the stellar mass+cluster-centric distance, at redshift $z=0$, 0.24 and 0.5, respectively.

On the other hand, we consider the SsHMR from the "simulation" point of view, meaning that we give predictions for the stellar mass at a given subhalo mass, and consider additional parameters which can be extracted from simulations (as opposed to data observables). We find that the scatter in the SsHMR is well correlated with the 3D distances to the host centres, but also that the orbital history of subhaloes is an even better predictor of the SsHMR than the instantaneous cluster-centric distance. Using for instance $x_{\mathrm{min}}$, the subhalo distance of closest approach to the host centre during its accretion history, allows to further reduce the scatter in the predicted satellite stellar mass distribution. We also give fitting functions for the stellar mass as a function of either only the subhalo mass, the subhalo mass and cluster-centric radius, and the subhalo mass and distance of closest approach.

Finally, we examine in detail the time co-evolution of the dark matter and stellar components of satellite galaxies, since their time of first accretion within $2\times R_{200}$. We find that, as in \citetalias{niemiec2019}, the evolution is dominated by the tidal stripping of dark matter subhaloes. However, the new implementation of sub-grid physics in TNG with respect to Illustris yields some modifications in the evolution of galaxies during their infall into their host clusters: (i) galaxies form on average less stars during infall in TNG, which is probably due to the new galactic wind implementation \citep{pillepich2018a}; (ii) at the same time, tidal forces appear to be more efficient at stripping dark matter in TNG, which could be a selection effect due to the low statistics in Illustris, but could also underline the importance of baryon processes at the heart of dark matter subhaloes, and the necessity to properly account for the joint evolution of baryons and dark matter.

We also examine the impact of different galaxy properties on the stellar and dark matter mass evolution to better understand the mechanisms that drive these evolutions, and potentially generate the scatter in the SsHMR. We show that stellar mass at accretion influences the amount of star formation and stripping during infall, more massive galaxies being more prone to intrinsic quenching, and having a deeper infall towards the cluster centre, leading to an increase in stripping. The galaxy compactness at accretion also plays a part in determining the evolutionary path of galaxies, and in a way that is independent of the orbital history: more extended galaxies are subject to a larger amount of stripping, but also form more stars in the first stage of their infall. Conversely, the galaxy compactness as measured at redshift $z=0$, appears to be a good indicator of the  orbital history of galaxies, more compact galaxies having a deeper infall, and therefore a larger loss of dark matter. Finally, we examine the role of preprocessing: we define preprocessed galaxies as the ones having reached their maximal subhalo mass prior to the start of their infall (we chose of conservative threshold of 1\,Gyr), and found that although this is a purely gravitational definition of preprocessing, it correlates well with the in-cluster vs pre-cluster quenching of the galaxies.

In this paper, we conducted an analysis based on measurables (mass, star-formation rate, etc) that are directly taken from the publicly available TNG simulation catalogues, extracted with friend-of-friend and \textsc{Subfind} algorithms. Although these quantities allow to gain some insights in the physical processes that lead to the observable properties of cluster galaxies, they are not necessarily completely equivalent to what can be observed. This is beyond the scope of this study, but we plan in a further analysis to connect the physical processes constrained here with more observationally measurable properties of satellite galaxies, such as their dark matter/stellar mass density profiles. This will also allow to plan how to best measure these quantities in observational analyses, for instance using galaxy-galaxy weak lensing in upcoming surveys such as Euclid-ESA mission \citep{euclidredbook}.
Predictions of the measurability of the tidal stripping in clusters should also be conducted in other simulations, including different implementations of the baryonic processes, or different dark matter candidates, to quantify the constraining power of observations on these processes, using satellite galaxies in clusters.

\section*{Acknowledgments}
The authors thank the anonymous referee for their comments that helped improve the clarity and the content of the paper.
We thank Arya Farahi for providing measurements of the host clusters concentrations and formations redshifts for the TNG300 simulation.
AN and MJ are supported by the United Kingdom Research and Innovation (UKRI) Future Leaders Fellowship `Using Cosmic Beasts to uncover the Nature of Dark Matter' (grant number MR/S017216/1). CG is supported by PRIN-MIUR
2017 WSCC32 ``Zooming into dark matter and proto-galaxies with massive
lensing clusters''.  CG acknowledges also support from the Italian Ministry
of Foreign Affairs and International Cooperation, Directorate General
for Country Promotion and CG from Italian Institute of Astrophysics under the grant "Bando PrIN 2019", PI: Viola Allevato.

\section*{Data Availability}
The data underlying this article is all based on the publicly available Illustris TNG simulation (Friend-of-Friend and \textsc{Subfind} catalogues, merger trees): \url{https://www.tng-project.org/data/}.

\bibliographystyle{mnras}
\bibliography{lensing.bib}

\bsp	
\label{lastpage}
\end{document}